\documentclass[11pt]{article}
\usepackage{amsfonts}
\usepackage{mathrsfs}

\usepackage[latin1]{inputenc}
\usepackage{float}
\usepackage{amsmath,amssymb}
\usepackage{latexsym}
\usepackage{epstopdf}
\usepackage{geometry}  
\usepackage[active]{srcltx}
\usepackage{graphicx}
\usepackage{epstopdf}
\usepackage{booktabs}
\usepackage{multirow}
\usepackage{siunitx}

\usepackage[
bookmarks=true,         
bookmarksnumbered=true, 
colorlinks=true, pdfstartview=FitV, linkcolor=blue, citecolor=blue,
urlcolor=blue]{hyperref}

\newtheorem {theorem}{Theorem}[section]

\newtheorem{definition}{Definition}[section]
\newtheorem{example}{Example}[section]
\newtheorem{lemma}{Lemma}[section]

\newtheorem{remark}{Remark}[section]

\newcommand{\bi}[1]{\mbox{\boldmath{$ #1 $}}}

\begin{document}

\title{Depth-based Weighted Jackknife Empirical Likelihood for Non-smooth $U$-structure Equations}
\author{Yongli Sang\textsuperscript{a}\thanks{CONTACT Yongli Sang. Email: yongli.sang@louisiana.edu}, Xin Dang\textsuperscript{b} and Yichuan Zhao\textsuperscript{c}}
\date{%
    \textsuperscript{a}Department of Mathematics, University of Louisiana at Lafayette, Lafayette, LA 70504, USA\\%
    \textsuperscript{b}Department of Mathematics, University of Mississippi, University, MS 38677, USA\\[2ex]%
     \textsuperscript{c}Department of Mathematics and Statistics, Georgia State University, Atlanta, GA 30303, USA\\[2ex]%
}

\maketitle

\begin{abstract}
In many applications, parameters of interest are estimated by solving some non-smooth estimating equations with $U$-statistic structure. Jackknife empirical likelihood (JEL) approach can solve this problem efficiently by reducing the computation complexity of the empirical likelihood (EL) method.  However, as EL, JEL suffers the sensitivity problem to outliers.  In this paper, we propose a weighted jackknife empirical likelihood (WJEL) to tackle the above limitation of JEL. The proposed WJEL tilts the JEL function by assigning smaller weights to outliers.  The asymptotic of the WJEL ratio statistic is derived. It converges in distribution to a multiple of a chi-square random variable. The multiplying constant depends on the weighting scheme. The self-normalized version of WJEL ratio does not require to know the constant and hence yields the standard chi-square distribution in the limit. Robustness of the proposed method is illustrated by simulation studies and one real data application.\\


\noindent  

\vskip.2cm 

\noindent {\bf Keywords:}
\noindent Weighted jackknife empirical likelihood; $U$-statistic structure equations; Robustness; Depth function\\


\end{abstract}

\section{Introduction}

The empirical likelihood (EL) method was first introduced by Owen (\cite{1988}, \cite{Owen1990}) and has been used heuristically  for constructing confidence regions.  It combines the effectiveness of likelihood and the reliability of nonparametric approach. On the computational side,  it involves a maximization of  the nonparametric likelihood supported on data subject to some constraints. If these constraints are linear, the computation of the EL method is particularly easy. However, when applied to some more complicated statistics such as $U$-statistics, it runs into serious computational difficulties. Many methods are proposed to overcome this computational difficulty,  for example, Wood, Do and Broom (\cite{Wood1996}) proposed a sequential linearization method by linearizing the nonlinear constraints. However, they did not provide the Wilks' theorem and stated that it was not easy to establish.  Jing, Yuan and Zhou (\cite{Jing2009}) proposed the jackknife empirical likelihood (JEL) approach. It transforms  the maximization problem of the EL with nonlinear constraints to the simple case of EL on the mean of  jackknife pseudo-values, which is very effective in handling one and two-sample $U$-statistics.  Since then, it has attracted strong interests in a wide range of fields due to its efficiency, and many papers are devoted to the investigation of the method, for example, Liu, Xia and Zhou (\cite{Liu2015}), Peng (\cite{Peng2012}), Feng and Peng (\cite{Feng2012}), Wang and Zhao (\cite{Zhao2016}), Wang, Peng and Qi (\cite{Wang2013}), Li, Xu and Zhou (\cite{Li2016}), Li, Peng and Qi (\cite{Li2011}), Sang, Dang and Zhao (\cite{Sang2017}) and so on. 
In many nonparametric and semiparametric approaches, such as the Gini correlation, quantile regression and rank regression, the parameters of interest are estimated by solving equations with $U$-statistic structure instead of directly by $U$-statistics.  Thus, the JEL in Jing, Yuan and Zhou (\cite{Jing2009}) can not be applied directly.  Li, Xu and Zhou (\cite{Li2016}) extended the JEL to the more complicated but more general situation. The Wilks' theorems are established even for the situation in which nuisance parameters are involved. 

As the EL method is sensitive to outliers and the EL confidence regions may be greatly lengthened in the directions of the outliers (Owen\cite{Owen2001}, Tsao and Zhou\cite{Tsao2001}), 
the JEL method with equation constraints is sensitive to outliers. That is, the JEL method is not robust. 
For the EL approach, a number of methods have been proposed to achieve robustness, see Wu (\cite{Wu2004}), Glenn and Zhao (\cite{Glenn2007}), Jiang $et$ $al.$ (\cite{Jiang2011}).  Those robust empirical likelihood (REL) methods tilt the EL function by assigning smaller weights to outliers, which yield a more robust estimator and confidence region.  
Jiang $et$ $al.$ (\cite{Jiang2011}) linked the depth-based weighted empirical likelihood (WEL) with general estimating equations and produced a robust estimation of parameters. They constructed weights based on a depth function although it is not the spatial depth as they claimed.  Data depth provides a centre-outward ordering of multi-dimensional data. Points deep inside the data are assigned with a high depth and those on the outskirts with a lower depth. In the literature, depth functions have been extensively studied,  for example, Mahalanobis depth (\cite{Rao1988}), simplicial depth (\cite{Liu1990}), half-space depth (\cite{Tukey75}), spatial depth (\cite{Serfling2002}) and projection depth (\cite{Zuo2003}). 
In this paper, we propose a weighted JEL (WJEL)  incorporating the depth-based weights in the JEL approach  to solve complicate problems with estimating $U$-statistic structure equations in order to gain robustness of the JEL procedure.   There is no smoothness assumption on the kernel function of U-statistic structure in the sample space. Rather, the smoothness on the parameter is required.  The asymptotic distribution of WJEL ratio is established.  It converges in distribution to a multiple of a chi-square random variable. The multiplying constant depends on the weighting scheme. The self-normalized version of WJEL ratio does not require to know the constant and hence yields the standard chi-square distribution in the limit. The proof of the limiting distribution of the WJEL is quite technically involved since the procedure has to deal with weak dependence of jackknife pseudo values and in the same time to deal with uneven weights. 
 
The remainder of the paper is organized as follows.  In Section 2,  we develop a weighted JEL (WJEL) method for estimating non-smooth $U$-statistic structure equations.   In Section 3, simulation studies are conducted to compare our WJEL methods with the JEL methods.  A real data analysis is illustrated in Section 4. Section 5 concludes the paper with a brief summary. All the proofs are reserved to the Appendix.

\section{Methodology}
\subsection{JEL with $U$-statistic structure equations}
Suppose that $\bi X_i$'s ($i=1, ..., n$) are independently distributed from an unknown distribution $F$ with a $r$-dimensional parameter $\bi \theta$. $\bi \theta$ can be estimated by solving the $U$-structure equations (Li, Xu and Zhou \cite{Li2016})
\begin{align*}
 W_{n, l}(\bi \theta)=[k!(n-k)!]/n!\sum_{1\leq i_1<..< i_k \leq n} H_l(\bi X_{i_1}, ..., \bi X_{i_k};\bi \theta)= 0,\;\;\;\; \mbox{for }\; l= 1, ..., r,
\end{align*}
where $ H_l(\cdot)$ is symmetric in the $\bi X$'s.  For each fixed $\bi \theta$, $W_{n, l}(\bi \theta)$ is a standard $U$-statistic with kernel $H_l(\cdot)$ for $l=1, ..., r$. Let $\bi H=(H_1, ..., H_r)^T$ and $\bi W_n=(W_{n, 1}, ..., W_{n, r})^{T}$.  
We define the jackknife pseudo-values by  
\begin{align} \label{jpv}
\hat{\bi V}_i(\bi \theta)=n \bi W_n-(n-1) \bi W^{(-i)}_{n-1},
\end{align}
where $\bi W^{(-i)}_{n-1}= \bi W_{n-1}(\bi X_1, ..., \bi X_{i-1}, \bi X_{i+1}, ..., \bi X_{n}; \bi \theta )$ is calculated on the sample of $n-1$ data values from the original data set after the $i^{th}$ observation is deleted. It has been proved that those jackknife pseudo-values are asymptotically independent (\cite{Shi1984}), and the average of those jackknife pseudo-values provides a unbiased and consistent estimator of $\mathbb{E}\bi H(\bi X_1,...,\bi X_k; \bi \theta)$. Therefore, the standard empirical likelihood can be established on those pseudo-values instead of the original observations $X_1,...,X_n$ as follows.  
The $U$-type empirical likelihood (UEL) at $\bi \theta$ is given by
\begin{align}\label{jel}
L(\bi \theta)=\max\left\{\prod_{i=1}^{n} p_i: \sum_{i=1}^{n}p_i=1, \sum_{i=1}^{n}p_i\hat{\bi V}_i(\bi \theta)=\bi 0\right\}.
\end{align}

Using Lagrange multipliers, when $\bi 0$ is in the convex hall of $\hat{\bi V}_i(\bi \theta), i=1, ..., n$, 
\begin{align*}
p_i=\frac{1}{n}\frac{1}{1+\bi \lambda^T(\bi\theta)\hat{\bi V}_i(\bi \theta)},
\end{align*}
where $\bi \lambda(\bi\theta)=(\lambda_1(\bi \theta), ..., \lambda_r(\bi \theta))^T$ are the Lagrange multipliers that satisfy
\begin{align*}
\frac{1}{n}\sum_{i=1}^{n}\frac{\hat{\bi V}_i(\bi \theta)}{1+\bi \lambda^T(\bi \theta)\hat{\bi V}_i(\bi \theta)}=\bi 0.
\end{align*}

Under some mild conditions listed in Li, Xu and Zhou (\cite{Li2016}), the Wilks' theorem holds for the $U$-type empirical likelihood ratio, that is, as $n \to \infty$, 
$$-2\log R(\bi \theta_0)=2 \sum_{i=1}^{n}\log\left \{1+\bi \lambda^T(\bi \theta_0)\hat{\bi V}_i( \bi \theta_0)\right \}   \stackrel{d}{\rightarrow} \chi^{2}_r, $$ 
where $\bi \theta_0$ is the true value of $\bi \theta$. 

In many nonparametric and semiparametric approaches, the parameters of interest are estimated by solving equations with $U$-statistic structure instead of directly by $U$-statistics.  

\begin{example}[Gini correlation]
Gini correlation, as an alternative measure of dependence, can be estimated by solving non-smooth and $U$-structured estimating functions (Sang, Dang and Zhao~\cite{Sang2017}). Specifically, 
suppose $X$ and $Y$  are  two non-degenerate random variables  with continuous marginal distribution functions $F$ and $G$, respectively, and a joint distribution function $P$, then two Gini correlations are defined as (Blitz and Brittain~\cite{BB64}, Schechtman and Yitzhaki~\cite{Yitzhaki2013})
\begin{align}\label{eqn:rho}
&\gamma_1:=\gamma(X,Y)=\frac{\mbox{cov}\big(X, G(Y)\big)}{\mbox{cov}\big(X,F(X)\big)} \;\;\;\text{and }\;
\gamma_2:=\gamma(Y, X)=\frac{\mbox{cov}\big(Y, F(X)\big)}{\mbox{cov}\big(Y,G(Y)\big)}.
\end{align}

Given an i.i.d. data set  $\mathcal{Z}=\{\bi Z_1, \bi Z_2, ..., \bi Z_n\}$ with $\bi Z_i=(X_i, Y_i)^T$, the two Gini correlations  can be estimated by a ratio of two $U$-statistics 
\begin{align*}
&\hat{\gamma}_1=\frac{U_1}{U_2} =  \frac{ 2/[n(n-1)]\sum_{1\le i<j \le n}h_1\big((X_i,Y_i),(X_j,Y_j)\big)}{2/[n(n-1)]\sum_{1\le i<j \le n}h_2\big((X_i,Y_i),(X_j,Y_j)\big)},\\
&\hat{\gamma}_2=\frac{U_3}{U_4} =  \frac{ 2/[n(n-1)]\sum_{1\le i<j \le n}h_1\big((Y_i,X_i),(Y_j,X_j)\big)}{2/[n(n-1)] \sum_{1\le i<j \le n}h_2\big((Y_i,X_i),(Y_j,X_j)\big)},
\end{align*}
where
$h_{1}\big((x_1, y_1), (x_2, y_2)\big)= 1/4[(x_1-x_2)I (y_1>y_2) +(x_2-x_1)I(y_2>y_1)]$ and  
$h_{2}\big((x_1, y_1), (x_2, y_2)\big)=1/4|x_1-x_2|$. Let $\bi \gamma=(\gamma_1, \gamma_2)^T$ and $\bi H=(H_1, H_2)^T$ with 
\begin{align*}
H_1\big((x_1,y_1), (x_2,y_2); \bi \gamma\big )=h_2\big((x_1,y_1), (x_2,y_2)\big) \gamma_1-h_1\big((x_1,y_1), (x_2,y_2)\big),\\
H_2\big((x_1,y_1), (x_2,y_2); \bi \gamma\big)=h_2\big((y_1,x_1), (y_2,x_2)\big) \gamma_2-h_1\big((y_1,x_1), (y_2,x_2)\big).
\end{align*} 
Then from Sang, Dang and Zhao (\cite{Sang2017}), we have the following $U$-structure equations,
\begin{align*}
 \bi W_{n}(\bi \gamma)= \frac{2}{n(n-1)}\sum_{1\leq i<j \leq n} \bi H(\bi Z_i, \bi Z_j;\bi \gamma)= \bi 0, \;\;\;\; \; 
\end{align*}
Note that $H_1$ and $H_2$ are non-smooth with respect to the sample space since they involve the indicator function in $h_1$ and the absolute value function in $h_2$. However, it is differentiable with respect to the parameter.  
\end{example}

%

\begin{example}[Gini index]
Gini index has been widely used in Economics for assessing distributional inequality of income or wealth (\cite{Gini09}).   It can be estimated by solving non-smooth and $U$-structured estimating functions (\cite{Zhao2016}).
Let $X$ and $X'$ be a independent pair  of random variables from $F$. Then the Gini index of $F$ can be defined as follows,
\begin{align*}
GI=\dfrac{\mathbb{E}|X-X'|}{2\mathbb{E}X}.
\end{align*}

Given an i.i.d. data set  $\mathcal{X}=\{X_1,  X_2, ..., X_n\}$, a natural estimator for the Gini index is  a ratio of two $U$-statistics with the kernels $h_1(x_1,x_2)=|x_1-x_2|$ and $h_2(x_1,x_2)=(x_1+x_2)$, 
\begin{align}
\widehat{GI}=\dfrac{U_1}{U_2}=\dfrac{{n \choose 2}^{-1}\sum_{1 \leq i <j \leq n}|X_i-X_j|}{{n \choose 2}^{-1}\sum_{1 \leq i <j \leq n}(X_i+X_j)}.
\end{align}

Let $H(x_1,x_2)=h_2(x_1,x_2)\times GI-h_1(x_1,x_2)$. The Gini index can be estimated by
\begin{align*}
 W_{n}(GI)= \frac{2}{n(n-1)}\sum_{1\leq i<j \leq n} H(X_i, X_j;GI)= 0,
\end{align*}
which is a $U$-structure equation. Clearly, $H(x_1, x_2; GI)$ is non-smooth with respect to $x_1$ or $x_2$, but is smooth with respect to the parameter $GI$. 
\end{example}

\begin{remark}
The parameters in Example 1 and 2 are estimated by the ratios of the $U$-statistics, and they are biased. Using the theorem on a function of U-statistics, the limiting normality of the estimators can be established, but the JEL approach can avoid estimating their asymptotic variances.  Secondly, the JEL approach performs better in finite samples, especially in small samples. This is demonstrated empirically in \cite{Sang2017}. The weighted JEL is proposed in this paper with a goal to improve the robustness of the JEL.    
\end{remark}

\subsection{Weighted JEL with $U$-statistic structure equations }
In order to reduce the influence of outliers, we propose a robust JEL  by defining a weighted JEL  as follows.
\begin{definition}
Suppose that $\bi X_i$ ($i=1, ..., n$) are independent distributed from an unknown distribution $F$ with a $r$-dimensional parameter $\bi \theta$. Assume that $p_i$ is the probability mass placed on $\bi X_i$. Given a weight vector $\bi \omega_n$ with $\sum_{i=1}^{n}\omega_{ni}=1$ and $\omega_{ni} \geq0$, the weighted jackknife empirical likelihood (WJEL)  for parameter $\bi \theta$ is then defined as
\begin{align}\label{WJEL}
\text{\mbox{WJEL}}(\bi \theta)=\sup\left \{\prod_{i=1}^{n}p^{n\omega_{ni}}_i: \sum_{i=1}^{n} p_i=1,  \sum_{i=1}^{n}  p_i\hat{\bi V}_i(\bi \theta)=\bi 0 \right\},
\end{align}
where $\hat{\bi V}_i(\bi \theta), i=1,...,n$ are the jackknife pseudo values defined in (\ref{jpv}).  
\end{definition}
\begin{remark}
For the equal weights $\omega_{ni}=\dfrac{1}{n}, i=1,2, ..., n$, the WJEL defined as (\ref{WJEL}) is reduced to the classical JEL in (\ref{jel}). 
\end{remark}

\begin{remark}
The parameter $\bi \theta$ in (\ref{WJEL}) is not directly related with the weight vector $\bi \omega_n$ that is given or specified. However, since the WJEL  is related with both the parameters and the weights, maximizing the jackknife empirical log-likelihood ratio brings an indirect connection between the parameter and weights. 
\end{remark} 

We defer the choice of $\bi \omega_n$ for robustness of the WJEL to the end of this section, but focus on the solution of (\ref{WJEL}) and its asymptotic property first.   We want to maximize $\prod_{i=1}^{n} p_i^{n\omega_{ni}}$ subject to restrictions 
\begin{align*}
p_i \geq 0,\;\;\; \sum_{i=1}^{n}p_i=1, \;\;\;  \sum_{i=1}^{n}p_i\hat{\bi V}_i(\bi \theta)=\bi 0.
\end{align*}
For any given $\bi \theta$, if $\bi 0$ is in the convex hull of points $\hat{\bi V}_1(\bi \theta), ..., \hat{\bi V}_n(\bi \theta)$, then a unique maximum exists and it can be found by Lagrange multipliers as follows,
\begin{align}\label{pi}
p_i=\dfrac{\omega_{ni}}{1+\bi \lambda^T \hat{\bi V}_i(\bi \theta)},
\end{align} 
where $\bi \lambda$ can be determined in terms of $\bi \theta$ by
\begin{align}\label{lambda}
\sum_{i=1}^{n}\dfrac{\omega_{ni}\hat{\bi V}_i(\bi \theta)}{1+\bi \lambda^T \hat{ \bi V}_i(
\bi \theta)}=0.
\end{align}
We can rewrite the WJEL function for $\bi \theta$ as 
\begin{align*}
L(\bi \theta)=\prod_{i=1}^n \left \{\dfrac{\omega_{ni}}{1+\bi \lambda^T \hat{\bi V}_i(\bi \theta)}\right \}^{n \omega_{ni}}.
\end{align*}
Note that the unrestricted empirical likelihood $\prod^n_{i=1}p^{n\omega_{ni}}_i$ is maximized at $p_i=\omega_{ni}, i=1,...,n$ because the Kullback-Leibler (KL) divergence $KL(\bi \omega_n|\bi p)$ defined as $-\sum_{i=1}^n \omega_{ni} \log(p_i/\omega_{ni})\geq 0$,  the equality if and only if $\bi \omega_{n}=\bi p$ (\cite{Kullback51}).   Then the corresponding  robust jackknife  empirical likelihood and robust jackknife  empirical log-likelihood ratio, respectively, are
\begin{align*}
R(\bi \theta)=\prod_{i=1}^{n}\left(\frac{p_i}{\omega_i}\right)^{n\omega_{ni}}= \prod_{i=1}^n \left \{\dfrac{1}{1+\bi \lambda^T \hat{\bi V}_i(\bi \theta)}\right \}^{n \omega_{ni}}
\end{align*}
and 
\begin{align}\label{jelr}
l(\bi \theta) =-2 \log R(\bi \theta)=2\sum_{i=1}^{n} n\omega_{ni} \log \{1+\bi \lambda^T  \hat{ \bi V}_i(\bi  \theta) \}.
\end{align}
From (\ref{jelr}), it is easy to see that the weight $n\omega_{n_i}$ is not assigned to the jackknife pseduo value $\hat{\bi V}_i(\theta)$ but to the empirical log-likelihood term $\log \{1+\bi \lambda^T  \hat{ \bi V}_i(\bi  \theta) \}$.  
We can minimize $l( \bi \theta)$ in (\ref{jelr}) to obtain an estimator $\tilde{\bi \theta}$ of $\bi \theta$, which is named as the WJEL estimator. We also have the following asymptotic result for the robust jackknife empirical log-likelihood ratio.


\begin{theorem}\label{aschi}
Let $\bi \theta_0$ be the true value of $\bi \theta$ and $c=\lim_{n \to \infty}\sum_{i=1}^n n \omega^2_{ni}$. Under some mild regularity conditions stated in the Appendix, the Wilks theorem holds for the $U$-type WJEL ratio,
\begin{align}
&l(\bi \theta_0) \stackrel{d}{\rightarrow} c\chi^2_r  \;\; \text{as}\;\; n\to \infty  \label{aschi1}\\
&\dfrac{l(\bi \theta_0)}{\sum_{i=1}^n n \omega^2_{ni}} \stackrel{d}{\rightarrow} \chi^2_r\;\; \text{as}\;\; n\to \infty.\label{aschi2}
\end{align}
\end{theorem}

The self-normalized result in (\ref{aschi2}) is more applicable since it does not require to know $c$ value. Its proof immediately follows from an application of Slutsky's Theorem to  (\ref{aschi1}), and hence we only provide a proof of (\ref{aschi1}) in the Appendix. 

The above procedure can also be adapted to handle nuisance parameters by  profiling the empirical likelihood. 
Write $ \bi \theta=(\bi \alpha^T,  \bi \beta^T)^T$, where $\bi \alpha \in \mathbb{R}^p$ is the parameter of interest and 
$\bi \beta \in \mathbb{R}^q$ $(p+q=r)$ is an unknown nuisance parameter. The profile WJEL ratio is defined as
\begin{align}\label{tiltebeta}
l(\bi \alpha)=l(\bi \alpha, \tilde{\bi \beta})= \min_{\bi \beta}l(\bi \alpha, \bi \beta).
\end{align}
That is, we minimize the WJEL ratio over the nuisance parameters for each fixed $\bi \alpha$.

\begin{theorem}\label{aschinuisance}
Under some mild regularity conditions stated in the Appendix, the Wilks' theorem holds for the $U$-type profile WJEL ratio and its  self-normalized version,
\begin{align}
&\bi l(\bi \alpha_0) \stackrel{d}{\rightarrow} c\chi^2_p \;\; \text{as}\;\; n\to \infty, \label{asprofile}\\ 
&\dfrac{l(\bi \alpha_0)}{\sum_{i=1}^n n \omega^2_{ni}} \stackrel{d}{\rightarrow} \chi^2_p\;\; \text{as}\;\; n\to \infty. \label{asprofile_self}
\end{align}
where  $\bi \alpha_0$ is the true value of the parameter of interest $\bi \alpha$ and $c$ is the same as in the Theorem \ref{aschi}. 
\end{theorem}

A proof for (\ref{asprofile}) of Theorem \ref{aschinuisance} is reserved in the Appendix.  The above results are obtained under a given weight vector $\bi \omega_n$.  In order to achieve the robustness with JEL, the weight $\omega_{ni}$ for an outlier $X_i$ should be small. For this propose, a proper weight scheme can be assigned by depth functions (Zuo and Serfling \cite{Zuo2000a}, and Dang, Serfling and Zhou \cite{Dang2009}). 
\subsection{Depth-based weights}
Depth functions play important roles in robust and nonparametric multivariate analysis and inference (Liu~\cite{Liu1999}, Zuo and Serfling~\cite{Zuo2000a}). Let $F$ be a probability distribution on $\mathbb{R}^d$.  An associated depth function $D(\bi x, F )$ provides a center-outward ordering of point $\bi x \in \mathbb{R}^d$, higher values representing higher ``centrality" of $\bi x$.  For a data set ${\cal X}_n = \bi (\bi X_1,...,\bi X_n)$ with the empirical distribution $F_n$, we will denote the sample version by $D(\bi x, F_n)$, which assigns points deep inside the data with a high depth and those on the outskirts with a lower depth. 

Among popular types of depth functions, we use the spatial depth for its nice properties in good balance between robustness and computational ease (Dang and Serfling~\cite{Dang2010}).  The spatial depth function is defined as 
 \begin{align}\label{depth}
 D(\bi x; F)= 1- \| \mathbb E_F \bi S(\bi x -\bi X) \|, 
 \end{align}
where $\bi S(\bi x)$ is the multivariate sign function with $\bi S(\bi x) = \bi x /\|\bi x\| $ if $\bi x \neq \bi 0$ and $\bi 0 $ if $\bi x =\bi 0$.   Accordingly, its sample counterpart is 
$$ D(\bi x, F_n) = 1- \left \|\frac{1}{n} \sum_{i=1}^n \frac{\bi x -\bi X_i}{\|\bi x -\bi X_i\|} \right\|. $$
It is easy to check that in the univariate case, the spatial depth is $D(x,F) =1-|2F(x)-1|$ and the sample spatial depth is $D(x,F_n) =1-|2F_n(x)-1|$ with the maximum spatial depth 1 at the median.  

Now we are ready to assign a weight to $\bi X_i$ by 
 \begin{align}\label{sw}
 \omega_{ni}=\dfrac{D(\bi X_i; F_n)}{\sum_{j=1}^n D(\bi X_j; F_n)}.
 \end{align}
 
 \begin{remark} \label{ksd}
When the data set is multimodal, it is suggested to use the kernalized spatial depth (KSD), which generalizes the  spatial  depth  via  a  positive  definite  kernel to capture  the  local  structure  of  the data cloud \cite{Chen09}. 
 \end{remark}
 
Note that Jiang {\em et al.} (\cite{Jiang2011}) used a different depth although they called it as the spatial depth function.  The depth they used is defined as $1/(1+\mathbb E_F \|\bi x -\bi X\|)$, which is not robust in terms of unbounded influence function and 0 breakdown point. Nevertheless, we can use Theorem 3.1 of Jiang {\em et al.} \cite{Jiang2011}, from which the constant $c$ in Theorem \ref{aschi} or Theorem \ref{aschinuisance} can be determined by  equation (\ref{cons:c}). 
\begin{theorem} \label{thm:depth}
If $\int D(\bi x; F) dF(\bi x)>0$ and $0 \leq D(\bi x; F) \leq 1$.  Then
 $\omega_{ni} \geq 0$, $\sum_{i=1}^n \omega_{ni}=1$, and as $n \to \infty$, 
 \begin{align}\label{cons:c}
 \sum_{i=1}^n n \omega^2_{ni}  \stackrel{a.s.}{\longrightarrow} c = \dfrac{\int D^2(\bi x; F) dF(\bi x)}{\big (\int D(\bi x; F) dF(\bi x)\big )^2}.
 \end{align}
 \end{theorem}
A direct application of the Jensen's inequality on a non-degenerate distribution $F$ proves that the spatial depth satisfies the conditions of  $\int D(\bi x; F) dF(\bi x)>0$ and $0 \leq D(\bi x; F) \leq 1$. 

\begin{remark}
The spatial-depth based weights provided in this section are data driven for the robustness purpose. This choice of weights may not satisfy the assumption of Theorem \ref{aschi} and Theorem \ref{aschinuisance}, in which the weights are required to be given and to be deterministic. However, the WJEL based on spatial-depth weights, (\ref{sw}) works well in the simulations studies, and definitely calls for a theoretic development  in the future research.  
\end{remark}

\section{Simulation}

In the first part of this section, a small simulation study is conducted to compare WJEL and JEL methods for inference of Gini correlation.

\begin{itemize}
\item Data are generated from a normal distribution $F_1$ with contaminating distribution $F_2$, where
\begin{align*}
F_1 \sim N\Big ( (0,0)^T, \bi \Sigma \Big ), \,\, F_2 \sim N\Big( (0,0)^T, \bi 4 \Sigma \Big)
\end{align*}
with  $\bf\Sigma=\begin{pmatrix}1 \;\;\;&\rho\\ \rho\;\;\;&1\end{pmatrix}$ and $\rho$ being the correlation coefficient.  Without loss of generality, we consider only cases of $\rho>0$ with $\rho=0.1, 0.5, 0.9$. 
We take two contamination levels: $0\%$ and $5\%$.
For each level, we generate 1000 samples of two different sample sizes ($n=20, 100$) from the mixture of $F_1$ and $F_2$. 
 For each simulated data set, $95\%$ confidence intervals  are calculated using different methods. The coverage probability and interval length can be computed from 1000 samples. Then we repeat this procedure 10 times. The average coverage probabilities and average lengths of confidence intervals as well as their standard deviations (in parenthesis) are presented in Table \ref{tab:normal}.

From Table \ref{tab:normal},  the WJEL method is more robust than the JEL method.  For the small size ($n=20$), the JEL performs worse than the WJEL even under the uncontaminated case. The WJEL keeps well the coverage probability while the JEL suffers a slight under-coverage problem.  For $n=100$ without outlier contamination, the JEL performs better than WJEL that produces slightly lower coverage probabilities.  
In the case of contamination, the WJEL always has higher coverage probabilities than JEL for all sample sizes. For a large sample size with $\rho=0.5$ or $\rho=0.9$, WJEL also yields shorter confidence intervals than JEL.  

\begin{table}[]
\caption{Coverage probabilities (standard deviations) and average lengths (standard deviations) of the Gini correlations' interval estimators from JEL and WJEL under mixture of bivariate normal and t(3) distributions.}
\center
\tiny
\scriptsize
\begin{tabular}{c|ll  cc   }
\hline \hline
Contamination levels&{$\rho$}&{Method}& {$n=20$}&  {$n=100$}   \\
&&&CovProb \ \ Length &CovProb  Length\\
\hline

&&JEL$\gamma_{1}$  &.929(.007) .840(.011)   
  & .951(.009) .338(.003) \\
 
 
 &&WJEL$\gamma_{1}$ &.948(.005)  1.02(.008)
 & .936(.012) .398(.003) \\
 
  
&$\rho=0.1$  &JEL$\gamma_{2}$  &.929(.005) .842(.006)   
  & .948(.009) .338(.003) \\
 
 
 &&WJEL$\gamma_{2}$ &.948(.005)  1.02(.009)
 & .932(.010) .398(.003) \\  \cline{2-5}
 

&&JEL$\gamma_{1}$  &.927(.007) .689(.006)   
  & .948(.010) .364(.003) \\
 
 
 &&WJEL$\gamma_{1}$ &.944(.005)  .839(.010)
& .934(.009) .325(.003) \\
 
  
$0\%$&$\rho=0.5$  &JEL$\gamma_{2}$  &.925(.007) .688(.008)   
  & .947(.006) .364(.002) \\
 
 
 &&WJEL$\gamma_{2}$ &.943(.005)  .840(.010)
 & .934(.010) .326(.003) \\ \cline{2-5}
 

&&JEL$\gamma_{1}$  &.921(.009) .247(.004)   
 & .945(.007) .099(.001) \\
 
 
 &&WJEL$\gamma_{1}$ &.942(.010)  .254(.004)
 & .928(.008) .091(.001) \\
 
  
&$\rho=0.9$  &JEL$\gamma_{2}$  &.915(.010) .246(.003)   
  & .946(.007) .099(.001) \\
 
 
 &&WJEL$\gamma_{2}$ &.938(.010)  .254(.003)
 & .928(.008) .091(.001) \\
 
\hline

&&JEL$\gamma_{1}$  &.923(.006) .865(.009)   
  & .945(.009) .363(.002) \\
 
 
 &&WJEL$\gamma_{1}$ &.946(.007)  1.05(.007)
 & .940(.010) .414(.003) \\
 

&$\rho=0.1$  &JEL$\gamma_{2}$  &.922(.009) .864(.009)   
  & .945(.010) .363(.002) \\
 
 
 &&WJEL$\gamma_{2}$ &.948(.008)  1.05(.007)
 & .940(.012) .414(.002) \\  \cline{2-5}
 

&&JEL$\gamma_{1}$  &.915(.008) .699(.008)   
  & .944(.007) .381(.002) \\
 
 
 &&WJEL$\gamma_{1}$ &.939(.007)  .854(.006)
& .939(.005) .343(.003) \\
 
  
$5\%$&$\rho=0.5$  &JEL$\gamma_{2}$  &.917(.010) .700(.008)   
  & .943(.006) .381(.001) \\
 
 
 &&WJEL$\gamma_{2}$ &.941(.007)  .849(.008)
 & .938(.006) .343(.002) \\  \cline{2-5}
 

&&JEL$\gamma_{1}$  &.916(.011) .253(.004)   
 & .941(.009) .104(.001) \\
 
 
 &&WJEL$\gamma_{1}$ &.943(.008)  .262(.005)
 & .934(.009) .096(.001) \\
 
  
&$\rho=0.9$  &JEL$\gamma_{2}$  &.917(.012) .254(.005)   
  & .939(.009) .104(.001) \\
 
 
 &&WJEL$\gamma_{2}$ &.941(.009)  .262(.005)
 & .934(.013) .096(.001) \\
 
   
\hline\hline
\end{tabular}
\label{tab:normal}
\end{table}

\item Data are generated from heavy-tailed  distributions. To be specific, we  generate 1000 samples of two different sample sizes ($n=20, 100$) from Kotz distribution with the scatter matrix $\bf\Sigma$ as before. The Kotz type distribution is a bivariate generalization of the Laplace distribution with the tail region fatness between that of the normal and $t$ distributions. The results based on 10 repetitions are presented in Table \ref{tab:t}.

\begin{table}[]
\caption{Coverage probabilities (standard deviations) and average lengths (standard deviations) of the Gini correlations' interval estimators from JEL and WJEL under Kotz distributions.}
\center
\scriptsize
\begin{tabular}{c|llcccc}
\hline \hline
Distribution &{$\rho$}&{Method}& {$n=20$}&  {$n=100$}   \\
&&&CovProb  Length &CovProb  Length\\
\hline

&&JEL$\gamma_{1}$  &.910(.010) .938(.008)   
  & .940(.008) .394(.004) \\
 
 
 &&WJEL$\gamma_{1}$ &.952(.006)  1.15(.009)
 & .955(.006) .475(.003) \\
 
  
&$\rho=0.1$  &JEL$\gamma_{2}$  &.909(.011) .936(.011)   
  & .942(.009) .393(.004) \\
 
 
&&WJEL$\gamma_{2}$ &.952(.006)  1.14(.008)
 & .955(.007) .475(.002) \\
 
\cline{2-5}

&&JEL$\gamma_{1}$  &.904(.010) .751(.010)   
  & .940(.007) .429(.003) \\
%
 
 &&WJEL$\gamma_{1}$ &.942(.007)  .931(.010)
& .953(.004) .398(.003) \\
 
  
 Kotz&$\rho=0.5$  &JEL$\gamma_{2}$  &.903(.011) .752(.009)   
  & .937(.007) .429(.003) \\
 
 
 &&WJEL$\gamma_{2}$ &.944(.009)  .930(.009)
 & .951(.004) .397(.003) \\
 
\cline{2-5}

&&JEL$\gamma_{1}$  &.897(.010) .272(.007)   
 & .938(.006) .118(.001) \\
 
 
&&WJEL$\gamma_{1}$ &.933(.007)  .285(.006)
 & .949(.007) .113(.001) \\
 
  
&$\rho=0.9$  &JEL$\gamma_{2}$  &.897(.014) .272(.008)   
  & .939(.009) .118(.001) \\
 
 
 &&WJEL$\gamma_{2}$ &.935(.008)  .285(.007)
 & .945(.010) .113(.001) \\
 

\hline\hline
\end{tabular}
\label{tab:t}

\end{table}

From Table \ref{tab:t}, under the heavy-tailed distributions, the JEL method suffers the under-coverage problem especially when the sample size is relatively small ($n=20$).  Compared with the JEL approach, the WJEL approach has better coverage probabilities which are very close to  the nominal  level with slightly larger average lengths of confidence intervals for both small and large sample sizes. 
Overall, the WJEL approach performs better than the JEL method in the heavy-tailed distributions. The WJEL overcomes the limitation of sensitivity of the JEL to outliers.  

%
\end{itemize}

The second part of the simulation studies is for comparing the JEL and WJEL methods when they are applied to  Gini index. We simulate data from asymmetric Pareto distributions, Pareto($\theta, \beta$), where $\theta$ is the scale parameter and $\beta$ is the shape parameter, respectively. 
Using the results of Gini mean difference in \cite{Zenga2004}, we have the true values of Gini index as follows, 
\begin{align*}
& \mbox{GI}(\mbox{Pareto}(\theta, \beta))=\frac{1}{2 \beta-1}, \;\;\;\;\text{for all $\theta>0$ and $\beta>1$}.
\end{align*}
The Gini index of Pareto distribution is independent of the scale parameter $\theta$. The number of samples and the number of repetitions are the same as before. The simulation results are reported in Table \ref{tab:index}.

From Table \ref{tab:index}, both the JEL and WJEL suffer the under-coverage problem, although the WJEL performs uniformly better than the JEL. The problem is more severe for the small sample size and small values of $\beta$. This is understandable since the smaller $\beta$ is, the heavier tail of the Pareto distribution is. The weighted JEL improves the JEL, but it does not reach the nominal level under the small sample size. As expected, the scale parameter $\theta$ has no impact on the  
inference of the Gini index. 

\begin{table}[]
\caption{Coverage probabilities (standard deviations) and average lengths (standard deviations) of interval estimators of Gini Index from JEL and WJEL methods under various Pareto distributions.}
\center
\scriptsize
\begin{tabular}{c|lccc}
\hline \hline

Pareto($\theta,\beta$) &{Method}& \multicolumn{1}{c}{$n=20$}&
 \multicolumn{1}{c}{$n=100$}\\ 
&&CovProb  Length &CovProb  Length\\
\hline

Pareto(1, 2)&JEL  &.704(.015) .301(.010)   
  & .815(.013) .171(.005) \\
 &WJEL &.715(.012)  .318(.010)
 & .831(.017) .198(.005) \\
 \hline


Pareto(1, 3)&JEL  &.781(.016) .204(.006)   
  & .878(.014) .101(.003) \\
 &WJEL &.796(.013)  .216(.006)
 & .887(.010) .119(.003) \\
 \hline
Pareto(4, 5) &JEL  &.819(.014) .116(.003)   
  & .907(.006) .051(.000) \\ 
 &WJEL &.825(.009)  .124(.003)
 & .914(.008) .062(.001) \\
 \hline

Pareto(1, 8)&JEL  &.842(.012) .069(.001)   
  & .920(.011) .029(.000) \\ 
 &WJEL &.849(.012)  .074(.001)
 & .930(.008) .036(.000) \\
 \hline
 
 Pareto(3, 8) &JEL  &.837(.012) .069(.002)   
  & .919(.010) .029(.000) \\ 
 &WJEL &.848(.013)  .074(.002)
 & .927(.011) .036(.000) \\
 \hline
 
Pareto(1, 15)&JEL  &.847(.006) .035(.001)   
  & .926(.005) .014(.000) \\ 
 &WJEL &.856(.009)  .038(.001)
 & .929(.004) .018(.000) \\
 \hline
 
Pareto(10, 15)&JEL  &.850(.012) .035(.001)   
  & .923(.008) .014(.000) \\ 
 &WJEL &.859(.013)  .038(.001)
 & .926(.007) .018(.000) \\
\hline\hline
\end{tabular}
\label{tab:index}

\end{table}

\section{Real data analysis}
For the purpose of illustration, we apply the proposed RJEL method to the gilgai survey data (Jiang $et$ $al.$\cite{Jiang2011}). The data set consists of 365 samples, which were taken at depths 0-10, 30-40 and 80-90 cm below the surface. Three features, pH, electrical conductivity (ec) in mS/cm and chloride content (cc) in ppm, are measured on a 1:5 soil:water extract from each sample. Without loss of generality, we consider the Gini correlations between electrical conductivity and chloride content at different depths. We use e00 (0-10 cm), e30 (30-40 cm), e80 (80-90 cm), and  c00 (0-10 cm), c30 (30-40 cm) and c80 (80-90 cm) to denote ec and cc at different depths, respectively . 

\begin{figure}[]
\caption{Density curves and boxplots of electrical conductivity  and chloride content at different depth levels.}
\centering
\label{densityfig}
\begin{tabular}{cc}
\includegraphics[width=2.9in,height=3.1in]{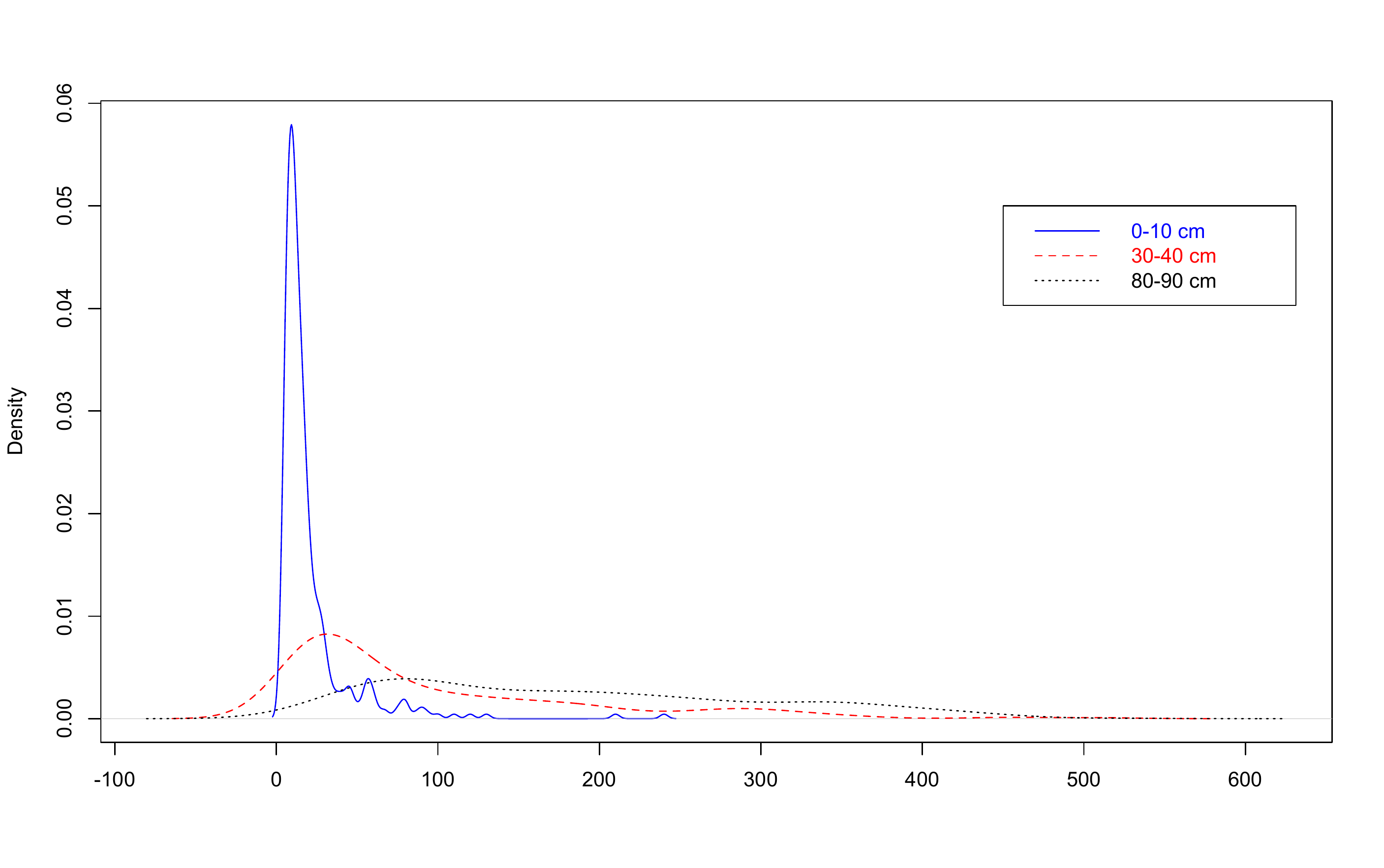} \vspace{-0.1in}&
\includegraphics[width=2.9in,height=3.1in]{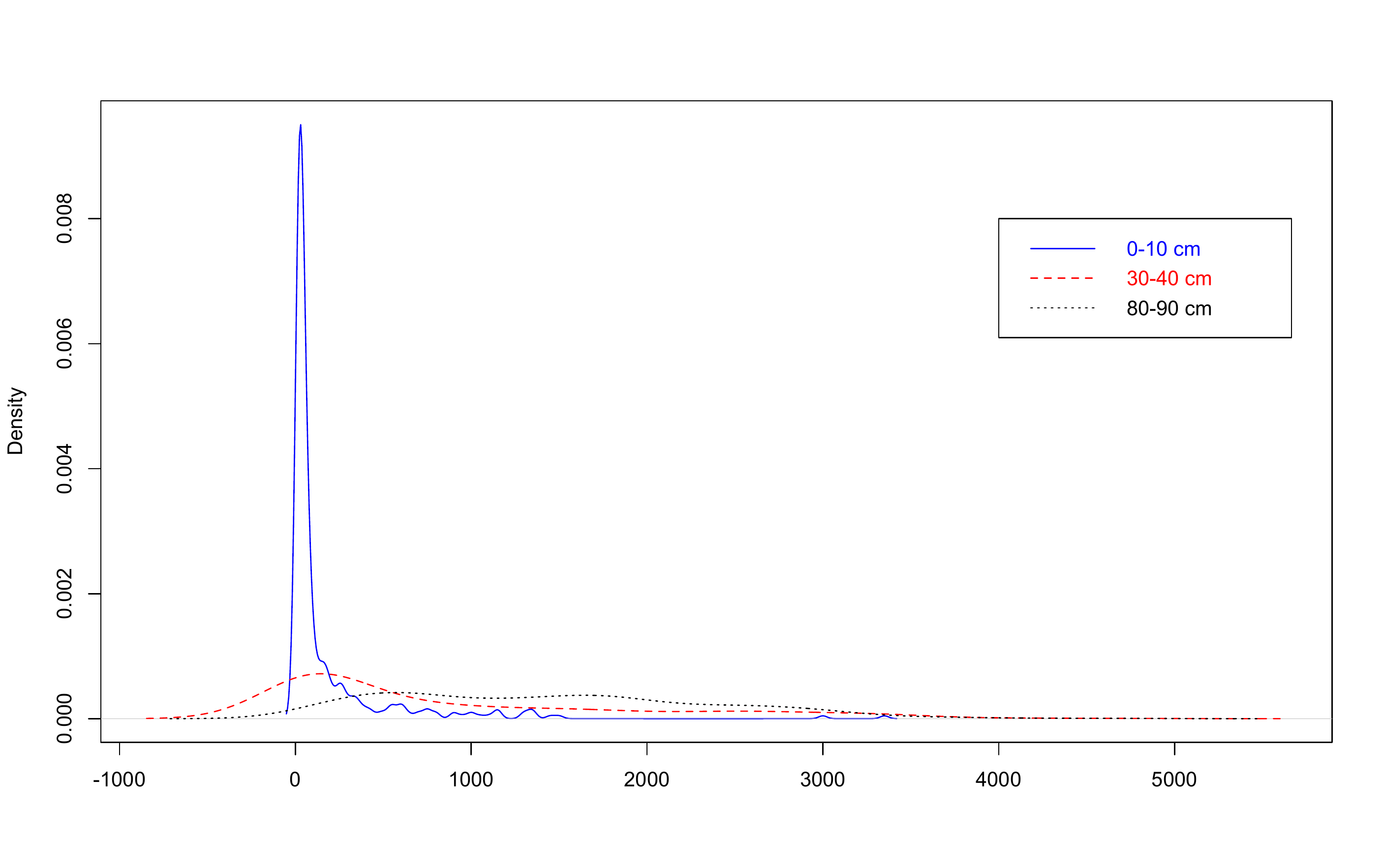} \vspace{-0.1in}\\
(a) density of ec &(b) density of cc \vspace{-0.1in}\\

\includegraphics[width=2.9in,height=3.1in ]{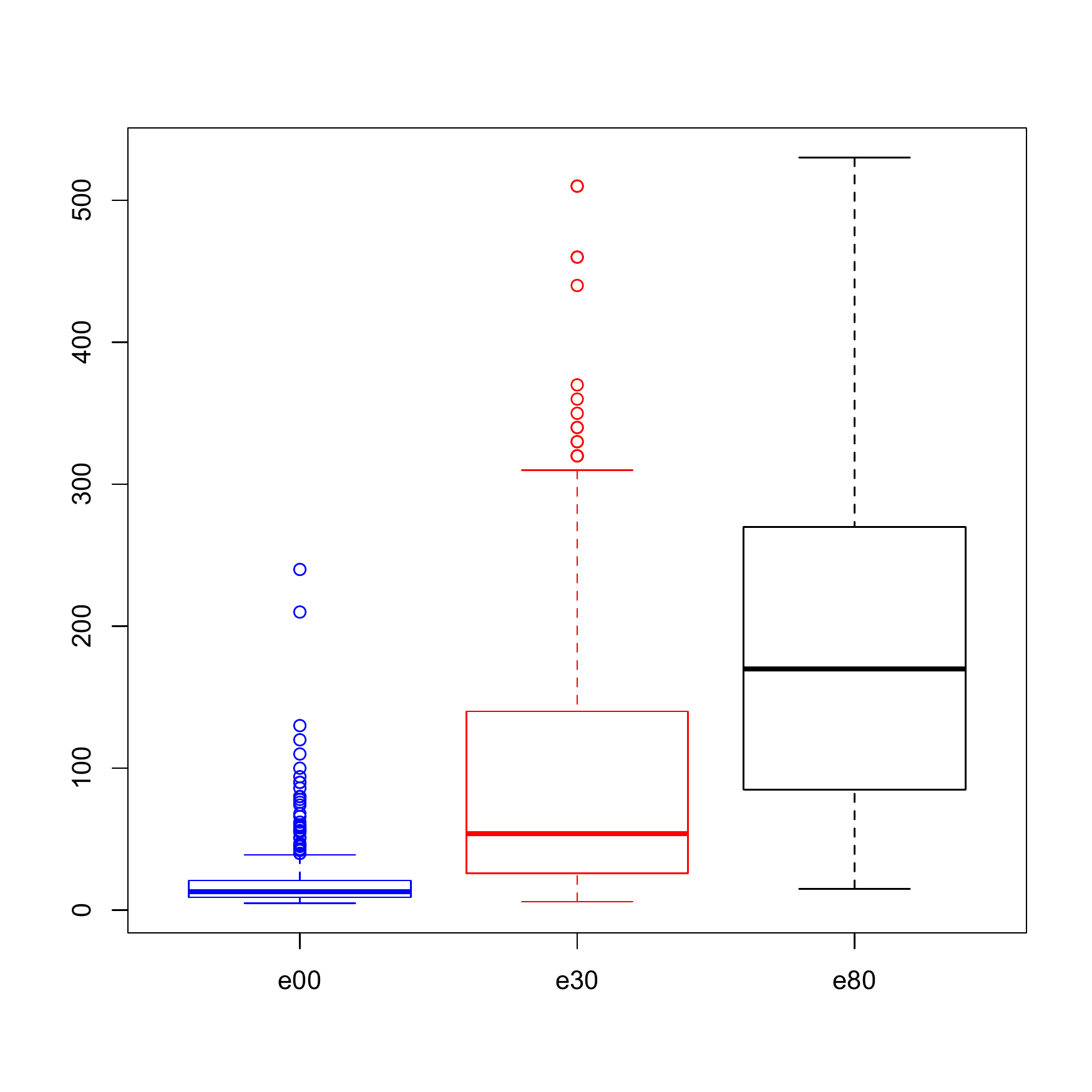} \vspace{-0.1in}&
\includegraphics[width=2.9in,height=3.1in]{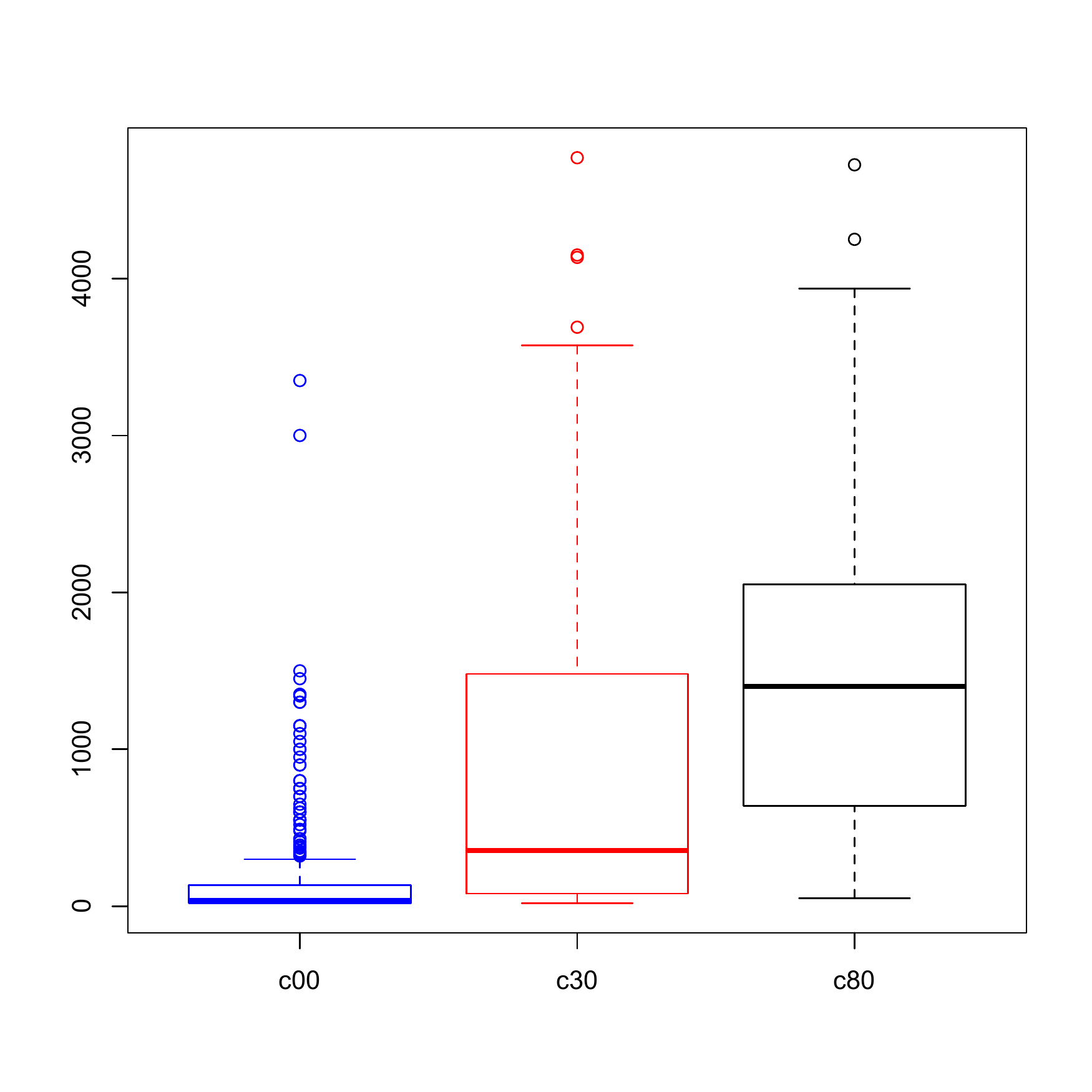} \vspace{-0.1in}\\
(c) boxplot of ec &(d) boxplot of  cc \vspace{-0.1in} \\
\end{tabular}
\label{fig:density}
\end{figure}

The density curves and boxplots of ec and cc at different depth levels are drawn in Figure \ref{fig:density}.  We observe that the distributions of each variable at different depths are quite different. The range and variation of each variable increase as the depth increases.  At the same depth, two features have a similar distribution although their scales are different.    Those distributions are positively skewed,   indicating that there are a quite number of outliers in two features at the 0-10 cm depth and few outliers at the 30-40cm depth. But e00 contains no outlier. 

The point estimates and confidence intervals for Gini correlations between three pairs (e00, c00), (e30, c30) and (e80, c80) are calculated and reported in Table \ref{tab4:test}.  We compare the RJEL with the other two methods, namely, JEL and VJ.  The VJ method is the inference method based on asymptotical normality with the asymptotical variance estimated by the jackknife method.  Correlations of ec and cc are significantly different at different depth levels. At the 30-40cm depth, the correlation is the highest between electrical conductivity and chloride content, while at the 80-90cm depth, the correlation between them is the lowest, decreasing from 0.97 to 0.73.



%

For (e00, c00), confidence intervals of $\gamma_1$ and $\gamma_2$ are disjoint by all three methods, indicating that  e00 and c00 is not exchangeable up to a linear transformation (Yitzhaki and Schechtman \cite{Yitzhaki2013}). With a large number of outliers presented in e00 and c00, performance of the JEL is largely degraded and is even worse than that of the VJ. The proposed RJEL overcomes the sensitivity of JEL to outliers and performs the best with the shortest confidence intervals. We keep more decimals in point estimates to see that the upper limits of the confidence intervals of the RJEL are slightly different with those point estimates. Indeed, that is one of appealing properties of the empirical likelihood approach: confidence intervals (regions) are entirely determined by data. 

For (e30, c30), the point estimates of $\gamma_1$ and $\gamma_2$ are very close and their confidence intervals are largely overlapped. Again, JEL performs the worst since both e30 and c30 contain few outliers. 

For (e80, c80), from the boxplots of Figure \ref{fig:density}, we know there are a couple of outliers in c80 but none in e80. Since $\gamma_1$ is defined as the covariance between e80 and the probability distribution of c80, the outliers in c80 do not impact JEL, and hence JEL performs better than RJEL. However, for inference of $\gamma_2$, JEL is affected by those outliers and RJEL is necessary. As a result, RJEL provides a much shorter confidence interval than the other two methods.

\begin{table}[]
\caption{Point estimates and $95\%$ confidence intervals for $\gamma_1$ and $\gamma_2$ for three pairs of (e00, c00), (e30, c30) and (e80, c80), respectively.}
\center
\scriptsize
\begin{tabular}{c|   cc  cc }
\hline \hline
Pairs&{Method}&{Point estimate}&{Confidence interval} & Interval length      \\
\hline
 &
JEL&
         & (.8311, .9166)& .0855 \\
&VJ&  $\hat{\gamma}_{1}=.8851901$
         & (.8499, .9204) & .0705\\   
          &RJEL&
         & (.8311, .8851904)& .0541 \\
\cline{2-5}

      (e00, c00)   &
JEL&
         & (.9274, .9663)& .0389 \\
 &VJ& $\hat{\gamma}_{2}=.9522348$
         & (.9361, .9684)& .0323 \\ 
         &RJEL&
         & (.9274, .9522351)& .0248 \\
\hline

 &
JEL&
         & (.9493, .9784) & .0291\\
&VJ& $\hat{\gamma}_{1}=.9700$& (.9565, .9836)& .0271\\
          &RJEL&
         & (.9631, .9899) & .0268\\
\cline{2-5}
      (e30, c30)   &
JEL&
         & (.9666, .9855)&.0189 \\
         &VJ&$\hat{\gamma}_{2}=.9798$&(.9712, .9885)& .0173\\
         &RJEL&
         & (.9754, .9931)&.0177 \\
 \hline

         &
JEL&
         & (.6414, .7812)& .1398 \\
&VJ& $\hat{\gamma}_{1}=.7279$&(.6716, .7843)& .1127\\
          &RJEL&
         & (.6991, .8432)& .1441 \\
\cline{2-5}
      (e80, c80)   &
JEL&
         & (.6842, .8198)& .1356 \\
         &VJ& $\hat{\gamma}_{2}=.7683408$& (.7135, .8232)& .1097\\
         &RJEL&
         & (.7403, .7683412)& .0280 \\
        
\hline\hline
\end{tabular}
\label{tab4:test}
\end{table}

\section{Conclusion}
In this paper, we have explored a robust JEL method for problems with $U$-statistic structure equations.  The RJEL is proposed to tilt the JEL function by assigning smaller weights to outliers with the weights being proportional to their spatial depth values.  Hence it is more robust than the regular JEL developed by Li, Xu and Zhou (\cite{Li2016}). Its robustness is demonstrated in the simulation and real data application on inference of Gini correlation. On the other hand, the asymptotic results of the robust JEL is the same as that of the robust EL (Jiang $et$ $al.$ (\cite{Jiang2011})), in which only general equation constraints are considered.  The proof of the asymptotic distribution of the RJEL  is quite technically involved since one has to deal with weak dependence of jackknife pseudo values and in the mean time to deal with unequal weights,  especially when the procedure involves the parameters of interest and nuisance parameters.


Continuations of this work could take the following directions: 
\begin{itemize}
\item In this paper, we use spatial depth function to assign weighs. How to assign  weight can be explored more in the further research.
\item Reduce the computation of  $U$-type profile empirical likelihood. Li, Peng and Qi (\cite{Li2011}) and Peng (\cite{Peng2012}) considered procedures based on a jackknife plug-in empirical likelihood to save the computation time. We may develop similar procedures to deal with $U$-structured empirical likelihood using the robust JEL.  
\end{itemize}

\section{Appendix}
Compared with  Li, Xu and Zhou (\cite{Li2016}), we need to deal with the uneven weight $\bi \omega_n$. For simplicity, we only prove Theorem \ref{aschinuisance} for $k=2$. The case for general $k$ and Theorem \ref{aschi} can be proved similarly. We adopt the notations from Li, Xu and Zhou (\cite{Li2016}) as below

\begin{enumerate}
\item $\bi \tau(\bi \beta)=\big (\tau_1(\bi\beta), ..., \tau_r(\bi\beta)\big)^T=\mathbb E \big (\bi H(X_1, X_2;\bi\alpha_0, \bi \beta)\big )$
\item $\hat{\bi V}_i(\bi\beta)=\hat{\bi V}_i(\bi \alpha_0, \bi\beta)$
\item $\bi W_n(\bi\beta)=\dfrac{1}{n}\sum_{i=1}^{n}\hat{\bi V}_i(\bi\beta)$
\item $S_n(\beta)=\sum_{i=1}^{n}\omega_{ni}\big (\hat{\bi V}_i(\bi \beta)- \bi \tau(\bi \beta)\big)\big(\hat{\bi V}_i(\bi \beta)-\bi \tau( \bi \beta)\big)^T$
\item $\bi \phi(x, \beta)=\big(\phi_1(x,  \bi \beta), ..., \phi_r(x,  \bi \beta)\big)^T=\mathbb E\big ( \bi H(x, X_1; \bi \alpha_0,  \bi \beta)\big )-\bi \tau( \bi\beta)$
\item  $\bi \psi(x, y,  \bi \beta)= \bi H(x,  y; \bi \alpha_0,  \bi \beta)-\bi \phi(x,  \bi \beta)-\bi \phi(y,  \bi\beta)-\bi \tau(\bi\beta)$
\item $\bi g(x, \bi \beta)=\big(g_1(x,  \bi\beta), ..., g_r(x,  \bi \beta)\big)^T=\bi \tau(\bi \beta)+2\bi \phi(x,  \bi\beta)$
\item $\sigma^2_l( \beta)=\mbox{var}\big(\phi_l(X_1,   \bi \beta)\big), l=1, ..., r$
\item $\sigma_{st}( \beta)=\mbox{cov}\big(\phi_s(X_1, \bi \beta), \phi_t(X_1, \bi \beta)\big), l=1, ..., r$
\item $\bi \Sigma^{(\bi \beta)}_{r\times r}$: the asymptotic variance-covariance matrix of $\sqrt{n}\big (\bi W_n( \bi \beta)-\bi \tau(\bi \beta)\big)$ with elements $4\sigma_{st}(\bi \beta), s,t= 1, ..., r$
\item $B(\bi \beta_0, \delta_n)=\{\bi \beta: \|\bi \beta-\bi \beta_0\|\leq \delta_n\},$ where $\{\delta_n\}$ is a sequence of non-negative real numbers converging to 0 as $n \to \infty$, and throughout this paper, we use some arbitrary $\delta_n$ unless otherwise specified 
\item $\Gamma_n(\bi\beta)=-\sum_{i=1}^n \omega_{ni} \log\big(1+\bi \lambda^T \hat{\bi V}_i(\bi \beta)\big)$
\item $\Gamma(\bi \beta)=-\mathbb E \{\log \big(1+\bi \xi^T \bi g(X_1,  \bi\beta)\big)\}$,
\end{enumerate}
where $\bi \lambda$ is determined by (\ref{lambda}) and $\bi \xi$ satisfies
\begin{align}\label{xi}
\mathbb E \left \{ \frac{\bi g(X_1, \beta)}{1+\bi \xi^T  \bi g(X_1,  \bi \beta)}\right \} =\bi 0.
\end{align}
Define the $(r+q) \times (r+q)$ matrix $V$
\begin{equation}\label{V}
V = 
\left(\begin{array}{cc}  V_{11} &V_{12}\\  V_{12}^T & 0 \end{array}\right),
\end{equation}
where
\begin{align*}
V_{11}=\bi \Sigma^{(\bi \beta_0)}_{r \times r}, \;\;\;V_{12}=-\dfrac{-\partial}{\partial \bi \beta^T}\mathbb E \big(\bi H(X_1, X_2; \bi \alpha_0, \bi\beta_0)\big).
\end{align*}

By the Hoeffding decomposition,
\begin{align*}
W_{n, l}(\bi \beta)= \tau_l(\bi \beta)+\frac{2}{n}\sum_{i=1}^n \phi_l(X_i, \bi \beta)+ 2/[n(n-1)]\sum_{i<j}\psi_{l}(X_i, X_j, \bi \beta).
\end{align*}
Simple calculations give
\begin{align}\label{hoeffding}
&\hat{V}_{il}(\bi \beta)\nonumber\\
&=\tau_l(\bi \beta)+2\phi_l(X_i, \bi \beta)+\frac{2}{n-1}\sum_{j=1, j\neq i}\psi_{l}(X_i, X_j, \bi \beta\nonumber\\
&- 2/[(n-1)(n-2)]\sum_{i_1<i_2, i_1\neq i, i_2 \neq i} \psi_l(X_{i1}, X_{i2}, \bi \beta)\nonumber\\
&:=g_l(X_i, \bi \beta)+R_{ni,l}(\bi\beta). 
\end{align}
Note that for each $\bi \beta \in B(\bi \beta_0, \bi\delta_n)$,
\begin{align*}
\mathbb E \big (R^2_{ni,l}(\bi \beta) \big)\leq C n^{-1}\mathbb E \big(\psi^2_l(X_{1}, X_{2}, \bi\beta)\big)+ C n^{-2}\mathbb E\big( \psi^2_l(X_{1}, X_{2}, \bi \beta) \big)\to 0,
\end{align*}
where $C$ is some generic constant. So, $R_{ni,l}=O_p(n^{-1/2})  \rightarrow 0$, which implies that $\hat{V}_{il}\stackrel{p}{\rightarrow}  g_l(X_i), i= 1, ..., n, l= 1, ..., r.$

We will need the following regularity conditions.

\noindent
(\textbf{C0}) The true parameter $(\bi \alpha_0, \bi \beta_0)$ is uniquely determined by $\mathbb E \big (\bi H(X_1, X_2; \bi \alpha_0, \bi \beta_0)\big)=0$;

\noindent
(\textbf{C1}) For $l= 1, ..., r$, when $(\bi \alpha, \bi \beta) \in \Theta$, $H_l(X_1, X_2; \bi \alpha, \bi \beta)$ is Borel measurable and uniformly bounded in $\mathbb{R}^{2d}\times \mathbb{R}^{r}$, where $X_1$ is a $d$ dimension random vector;

\noindent
(\textbf{C2}) For $\bi \beta \in B(\bi \beta_0, \delta_n)$, $\bi g(x, \bi \beta)$ is twice continuously differentiable with respective to $\bi \beta$;
$\mathbb E \{\bi g(X_1, \bi \beta)/[1+\bi \xi^T \bi g(X_1, \bi \beta)]\}$ is continuously differentiable in $\bi\beta \in B(\bi \beta_0, \delta_n)$ and $\bi \xi \in N_{\bi \xi}=\{\bi \xi: |\bi \xi| \leq \varepsilon_n\}$ (here $ \{\varepsilon_n\}$ is another sequence of non-negative real numbers converging to 0, as $n \to \infty$).  $\mathbb E \{ \bi g(X_1, \bi \beta)\bi g^T(X_1, \bi \beta)/[1+\bi \xi^T \bi g(X_1, \bi \beta)]^2\}$ is uniformly continuous in $\bi \beta \in B(\bi \beta_0, \delta_n)$ and $\bi \xi \in N_{\bi \xi}$, respectively;

\noindent
(\textbf{C3}) $\tilde{\bi \beta}$ defined by (\ref{tiltebeta}) converges to $\bi \beta_0$ in probability;

\noindent
(\textbf{C4}) The matrix $V_{11}$ defined in (\ref{V}) is positive definite;

\noindent
(\textbf{C5}) For $l= 1, ..., r$,
\begin{align*}
\sup_{\| \bi\beta -\bi\beta_0 \|\le \delta_n}  \dfrac{\|W_{n, l}(\bi \alpha_0, \bi \beta)-W_{n, l}(\bi \alpha_0, \bi \beta_0)-\tau_{l}( \bi \beta)\|}{n^{-3/2}+\|\bi \beta-\bi \beta_{0}\|}\stackrel{P}{\rightarrow} 0;
\end{align*}

\noindent
(\textbf{C6}) The weight vector $\bi \omega_n= (\omega_{n1}, \omega_{n2}, ..., \omega_{nn})^T$, $\omega_{ni}> 0, \sum_{i=1}^n \omega_{ni}=1$, satisfies that $\lim_{n \to \infty}\sum_{i=1}^n n \omega^2_{ni}=c$, where $c$ is a finite positive number. 

\begin{remark}
(i) Conditions \textbf{C0-C5} are the same as the conditions in Li, Xu and Zhou\cite{Li2016}. (ii) When all parameters are of interest, the kernel functions $H(\cdot)$  is not required for bounded.  (ii) \textbf{C6} can be satisfied by a wide range of depth functions. These conditions can be easily proved or found in literature for particular choices of $\bi H$.
\end{remark}

\noindent\textbf{PROOF OF THEOREM \ref{aschinuisance}.}
We first provide some useful lemmas.
\begin{lemma} [Hoeffding, 1948]\label{Hoeffding}
Under \textbf{C1}, we have 
\begin{align}
\sqrt{n}(\bi W_n(\bi \beta)- \bi \tau(\bi \beta))\stackrel{d}{\rightarrow} N(\bi 0,  \bi \Sigma^{( \bi \beta)}_{r 
\times r}), \; \text{as} \;\; n\to \infty
\end{align}
for each $\bi \beta \in B(\bi \beta_0, \delta_n)$.
\end{lemma}

\begin{lemma}\label{convexhull}
Under conditions \textbf{C1, C2, C4, C6}, with probability tending to one as $n \to \infty$, the zero vector is contained in the interior of the convex hull of    
$\left\{\omega_{n1}\hat{V}_1(\beta), ..., \omega_{nn}\hat{V}_n(\beta)\right\}$ for each $\bi \beta \in B(\bi \beta_0, \delta_n)$.
\end{lemma}
\textbf{Proof.}
From Lemma 2 of Li, Xu and Zhou (\cite{Li2016}), with probability tending to one as $n \to \infty$, the zero vector is contained in the interior of the convex hull of  $\left\{\hat{V}_1(\bi \beta), ...,\hat{V}_n(\bi \beta)\right\}$ for each $\bi \beta \in B(\bi \beta_0, \delta_n)$. As a result, the zero vector is contained in the interior of the convex hull of  $\left\{\omega_{n1}\hat{V}_1(\bi \beta), ..., \omega_{nn}\hat{V}_n(\bi \beta)\right\}$ for each $\bi \beta \in B(\bi \beta_0, \delta_n)$ since each $\omega_{ni}$ is positive.

\begin{remark}
Lemma \ref{convexhull} guarantees that 
 (\ref{lambda}) has a  solution $\bi \lambda$.
\end{remark}

\begin{lemma}\label{S}
Under \textbf{C1, C6}, we have $S_n(\bi \beta)=\bi \Sigma^{(\bi \beta_0)}_{r \times r}+o(1)$, a.s., where $o(\cdot)$ holds uniformly for $\bi \beta \in B(\bi \beta_0, \delta_n).$
\end{lemma}
\textbf{Proof.}
First, consider the diagonal elements of $S_n(\bi \beta)$. Let $S_{nl}(\bi \beta)$ denote the $l^{th}$
diagonal element of $S_n(\bi \beta)$, and $\hat{V}_{il}(\bi \beta)$  and $\mbox{W}_{nl}(\bi \beta)$ denote the $l^{th}$ component of $\hat{\bi V}_{i}(\bi \beta)$ and $\mbox{\bi W}_{n}(\bi \beta),$ respectively.

\begin{align}\label{expsdia}
&\mathbb{E} (S_{nl}(\bi \beta))\nonumber\\
&=\mathbb{E} \left \{\sum_{i=1}^n\omega_{ni}[\hat{V}_{il}(\bi \beta)-\tau_l(\bi \beta)]^2 \right \}\nonumber\\
&=\mathbb{E}\left \{\sum_{i=1}^n\omega_{ni}[\hat{\bi V}_{il}(\bi \beta)-\tau_l(\bi \beta)+\mbox{W}_{nl}(\bi \beta)-\mbox{W}_{nl}(\bi \beta)]^2\right \}\nonumber\\
&=\mathbb{E}[\mbox{W}_{nl}(\bi \beta)-\tau_{l}(\bi \beta)]^2+\sum_{i=1}^n\omega_{ni}\mathbb{E}[\hat{V}_{il}(\bi \beta)-\mbox{W}_{nl}(\bi \beta)]^2\nonumber 
\end{align}
\begin{align}
&+2\sum_{i=1}^n\omega_{ni}\mathbb{E}\left \{[\mbox{W}_{nl}(\bi \beta)-\tau_l(\bi \beta)][\hat{V}_{il}(\bi \beta)-\mbox{W}_{nl}(\bi \beta)] \right \}\nonumber\\
&=\mathbb{E}[\mbox{W}_{nl}(\bi \beta)-\tau_{l}(\bi \beta)]^2
+\mathbb{E}\left \{\frac{1}{n}\sum_{i=1}^n[\hat{V}_{il}(\bi \beta)-\mbox{W}_{nl}(\bi \beta)]^2\right \}\nonumber \\
&+2\mathbb{E}\left \{ [\mbox{W}_{nl}(\bi \beta)-\tau_l(\bi \beta)][\frac{1}{n}\sum_{i=1}^n(\hat{V}_{il}(\bi \beta)-\mbox{W}_{nl}(\bi \beta))]\right \}\nonumber\\
&=\mbox{var}(\mbox{W}_{nl}(\bi \beta))+(n-1)\mathbb{E} (\widehat{\mbox{var}}(Jack))\nonumber\\
&=\mbox{var}(\mbox{W}_{nl}(\bi \beta))+(n-1)\mbox{var}(\mbox{W}_{nl}(\bi\beta))\nonumber \\
&=4\sigma^2_{l}(\bi\beta),
\end{align}
which is the $l^{th}$ diagonal element of $\bi \Sigma^{(\bi \beta)}_{r \times r}$. In the above equations, $\widehat{\mbox{var}}(Jack)$  is the jackknife estimator of $\mbox{var}(\mbox{W}_{nl}(\bi \beta))$ and is defined (Lee\cite{Lee1990}, page 218)
 \begin{align*}
 \widehat{\mbox{var}}(\mbox{Jack})=1/[n(n-1)]\sum_{i=1}^n[\hat{V}_{il}(\bi \beta)-\mbox{W}_{nl}(\bi \beta)]^2.
  \end{align*}
By Shi (\cite{Shi1984}) and Shao and Tu (\cite{Shao1996}), for each fixed $\bi \beta$, $\hat{V}_{il}(\bi \beta)$'s are asymptotically independent. Applying the strong law of weighted sums of independent random samples in  Choi and Sung (\cite{Choi1987}) to $S_{nl}(\bi \beta)$, we have 
\begin{align}\label{sdia}
S_{nl}(\bi \beta)=\mathbb{E}S_{nl}(\bi \beta)+o(1), \ \text{a.s}.
\end{align}

Moreover,
\begin{align}\label{difdia}
&|\sigma^2_{l}(\bi \beta)-\sigma^2_{l}(\bi \beta_0)|=|\frac{\partial}{\partial \bi \beta^T}\mbox{var}(\phi_l(X_1, \bi \beta^{*}))(\bi\beta-\bi\beta_0)|\nonumber \\
&=|\frac{\partial}{\partial \bi \beta^T}\mathbb{E}(\phi^2_l(X_1, \bi\beta^{*}))(\bi\beta-\bi\beta_0)|\nonumber \\
&\leq C \|\bi\beta-\bi\beta_0\| \to 0,  
\end{align}
where $\bi\beta^{*}$ lies between $\bi\beta$ and $\bi\beta_0$, and C is a generic constant hereafter. 

In the second part,  we consider the off-diagonal elements of  $S_n(\bi\beta)$. Let $S_{n(ij)}(\bi\beta)$ denote the $ij^{th}$
off-diagonal element of $S_n(\bi\beta)$, $i, j=1,...,r$.

\begin{align}\label{expsoff}
&\mathbb{E}(S_{n(ts)}(\bi \beta))\nonumber\\
&=\mathbb{E}\left (\sum_{i=1}^n\omega_{ni}\big (\hat{V}_{it}(\bi \beta)-\tau_l(\bi \beta)\big)\big(\hat{V}_{is}(\bi \beta)-\tau_l(\bi\beta)\big)\right)\nonumber\\
&=\mathbb{E}[\sum_{i=1}^n\omega_{ni}\big(\hat{V}_{it}(\bi \beta)-\tau_s(\bi \beta)+\mbox{W}_{nt}(\bi\beta)-\mbox{W}_{nt}(\bi\beta)\big)\big(\hat{V}_{is}(\bi\beta)-\tau_s(\bi\beta)\nonumber\\ 
&+\mbox{W}_{ns}(\bi\beta)-\mbox{W}_{ns}(\bi\beta)\big)]\nonumber\\
&=\mathbb{E}[(\mbox{W}_{ns}(\bi \beta)-\tau_{s}(\bi\beta))(\mbox{W}_{nt}(\bi \beta)-\tau_{t}(\bi\beta))]\nonumber\\
&+\sum_{i=1}^n\omega_{ni}\mathbb{E}[(\hat{V}_{is}(\bi \beta)-\mbox{W}_{ns}(\bi \beta))(\hat{V}_{it}(\bi \beta)-\mbox{W}_{nt}(\bi \beta))]\nonumber\\
&+2\sum_{i=1}^n\omega_{ni}\mathbb{E}[(\mbox{W}_{ns}(\bi \beta)-\tau_s(\bi \beta))(\hat{V}_{it}(\bi \beta)-\mbox{W}_{nt}(\bi \beta))]\nonumber\\
&=\mathbb{E}[(\mbox{W}_{ns}(\bi \beta)-\tau_{s}(\bi \beta))(\mbox{W}_{nt}(\bi\beta)-\tau_{t}(\bi\beta))]\nonumber
\end{align}
\begin{align}
&+\mathbb{E}[\frac{1}{n}\sum_{i=1}^n(\hat{V}_{is}(\bi \beta)-\mbox{W}_{ns}(\bi \beta))(\hat{V}_{it}(\bi \beta)-\mbox{W}_{nt}(\bi \beta))]\nonumber\\
&+2\mathbb{E}[(\mbox{W}_{nt}(\bi \beta)-\tau_t(\bi \beta))(\frac{1}{n}\sum_{i=1}^n(\hat{V}_{is}(\bi \beta)-\mbox{W}_{ns}(\bi \beta)))]\nonumber\\
&=\mbox{cov}(\mbox{W}_{ns}(\bi \beta), \mbox{W}_{nt}(\bi \beta) )+(n-1)\mathbb{E}\widehat{\mbox{cov}}_{s,t}(Jack)\nonumber\\
&=\mbox{cov}(\mbox{W}_{ns}(\bi \beta), \mbox{W}_{nt}(\bi \beta) )+(n-1)\mbox{cov}(\mbox{W}_{ns}(\bi \beta), \mbox{W}_{nt}(\bi \beta) )\nonumber\\
&=4\sigma_{st}(\bi \beta),
\end{align}
which is the $st^{th}$ off-diagonal element of $\Sigma^{(\bi \beta)}_{r \times r}$. $\widehat{\mbox{cov}}_{s,t}(Jack)=1/n\sum_{i=1}^n(\hat{V}_{is}(\bi \beta)-\mbox{W}_{ns}(\bi \beta))(\hat{V}_{it}(\bi \beta)-\mbox{W}_{nt}(\bi \beta))$ is the jackknife estimator of $\mbox{cov}(\mbox{W}_{ns}(\bi \beta), \mbox{W}_{nt}(\bi \beta) )$.

Analogous to (\ref{sdia}), utilizing the strong law of weighted average on $S_{nts}(\bi \beta)$, we can show that
\begin{align}\label{soff}
S_{n(ts)}(\bi \beta)=\mathbb{E}(S_{n(ts)}(\bi \beta))+o(1).
\end{align}

Similar to (\ref{difdia}),
\begin{align}\label{difoff}
&|\sigma_{st}(\bi \beta)-\sigma_{st}(\bi \beta_0)|\nonumber \\
&=\Big |\frac{\partial}{\partial \bi \beta^T}\{\mathbb{E}[\phi_s(X_1,  \bi \beta^{*})\phi_t(X_1, \bi \beta^{*})]-\mathbb{E}[\phi_s(X_1, \bi \beta_0)\phi_t(X_1, \bi \beta_0)]\}(\bi \beta-\bi\beta^{*})\Big |\nonumber \\
&\leq C \|\bi\beta-\bi\beta_0\| \to 0,
\end{align}
where $\bi\beta^{*}$ lies between $\bi\beta$ and $\bi\beta_0$.
\begin{lemma}
Under \textbf{C1, C4} and \textbf{C6}, for $\tilde {\bi \beta}$ in (\ref{tiltebeta}), we have
\begin{align*}
\tilde{\bi \beta}=\arg\max_{\bi \beta}\Gamma_n(\bi \beta),\;\; \bi \beta_0=\arg\max_{\bi \beta}\Gamma(\bi\beta).
\end{align*}
\end{lemma}
The proof can be found in Li, Xu and Zhou (\cite{Li2016}) and Molanes Lopez, Van Keilegom and Veraverbeke (\cite{Molanes2009}).
\begin{lemma}\label{||lambda||}
For $\bi \lambda=\bi \lambda(\bi \beta)$ and $\bi\xi=\bi \xi(\bi\beta)$ satisfying (\ref{lambda}) and (\ref{xi}), under \textbf{C1-C6}, we have 
\begin{align}\label{order-lambda}
\|\bi \lambda\|=O_{p}(n^{-1/2})+O_{p}(\|\bi \beta-\bi\beta_0\|)
\end{align}
and $\bi\lambda-\bi\xi=O_p(n^{-1/2})$, uniformly for $\bi\beta \in B(\bi\beta_0, \delta_n)$. Moreover,
\begin{align*}
\bi \lambda=S^{-1}(\bi \beta)\sum_{i=1}^n \omega_{ni}\hat{\bi V}_i(\bi \beta)+o_p(n^{-1/2})
\end{align*}
uniformly for $\beta \in B(\beta_0, \delta_n)$ with $\delta_n=O_p(n^{-1/2})$.
\end{lemma}
\textbf{Proof.} 
By condition \textbf{C5},
\begin{align}\label{weight-sum}
&\sum_{i=1}^n \omega_{ni} \hat{\bi V}_i(\bi \beta)\nonumber \\
&=\sum_{i=1}^n \omega_{ni} \hat{\bi V}_i(\bi \beta_0)+\sum_{i=1}^n \omega_{ni}[ \hat{\bi V}_i(\bi \beta)-\hat{\bi V}_i(\bi \beta_0)]\nonumber \\
&=\sum_{i=1}^n \omega_{ni} \hat{\bi V}_i(\bi \beta_0)\nonumber\\
&+\sum_{i=1}^n \omega_{ni}[n \bi W_n(\bi \beta)-(n-1) \bi W^{(-i)}_{n-1}(\bi \beta)-n \bi W_n(\bi \beta_0)+(n-1) \bi W^{(-i)}_{n-1}(\bi \beta_0)]\nonumber \\
&=\sum_{i=1}^n \omega_{ni} \hat{\bi V}_i(\bi \beta_0)\nonumber 
\end{align}
\begin{align}
&+\sum_{i=1}^n \omega_{ni}\left \{n[\bi W_n(\bi \beta)- \bi W_n(\bi \beta_0)]-(n-1) [\bi W^{(-i)}_{n-1}(\bi \beta)-\bi W^{(-i)}_{n-1}(\bi \beta_0)] \right \}\nonumber \\
&=\sum_{i=1}^n \omega_{ni} \hat{\bi V}_i(\bi \beta_0)+\mathbb E  \bi H(X_1, X_2; \bi\alpha_0, \bi \beta)+o_p(\|\bi \beta-\bi \beta_0\|) +o_p(n^{-1/2})\nonumber \\
&=\sum_{i=1}^n \omega_{ni} \hat{\bi V}_i(\bi\beta_0)+\frac{\partial}{\partial \bi\beta^T}\mathbb E \bi H(X_1, X_2; \bi\alpha_0, \bi\beta^{*})(\bi\beta-\bi\beta_0)+o_p(n^{-1/2}) \nonumber \\
&+o_p(\|\bi\beta-\bi\beta_0\|),  
\end{align}
uniformly for $\bi\beta \in B(\bi\beta_0, \delta_n)$, where each element of $\bi\beta^{*}$ lies between the corresponding one of $\bi\beta$ and $\bi\beta_0$. 
By Lemma A.1 of Jiang $et$ $al.$ (\cite{Jiang2011}), the first term is $O_p(\sqrt{\sum_{i=1}^n \omega^2_{ni}})=O_p(n^{-1/2})$ by \textbf{C6}, and as $\frac{\partial}{\partial \bi\beta^T}\mathbb E\big ( \bi H(X_1, X_2; \bi\alpha_0, \bi\beta)\big)$ are uniformly continuous in $\bi\beta \in B(\bi\beta_0, \delta_n)$, we  have 
\begin{align}\label{weightaverage}
\sum_{i=1}^n \omega_{ni} \hat{\bi V}_i(\bi\beta)=O_p(n^{-1/2})+O_p(\|\beta-\beta_0\|),
\end{align}
uniformly for $\bi\beta \in B(\bi \beta_0, \delta_n)$.  Now, set $\bi \lambda=\rho \bi u$, where $\rho \geq 0$ and $\|\bi u\|=1$. It follows from (\ref{lambda}) that 
\begin{align}
&0=\|\sum_{i=1}^n \frac{\omega_{ni}\hat{\bi V}_i(\bi \beta)}{1+\bi \lambda^T\hat{\bi V}_i(\bi \beta )}\|\nonumber \\
&=\|\sum_{i=1}^n \omega_{ni} \hat{\bi V}_i(\bi \beta)-\bi \lambda \sum_{i=1}^n \frac{\omega_{ni} \hat{\bi V}_i(\beta)\hat{\bi V}^{T}_i(\bi \beta)}{1+\bi \lambda^T\hat{\bi V}_i(\bi \beta )} \|\nonumber\\
&\geq \rho \bi u^T \sum_{i=1}^n  \frac{\omega_{ni} \hat{\bi V}_i(\bi \beta)\hat{\bi V}^{T}_i(\bi \beta)}{1+\rho \bi u^T \hat{\bi V}_i(\bi \beta )}u-|\sum_{i=1}^n\omega_{ni}\bi u^T \hat{\bi V}_i(\bi \beta)|\nonumber \\
&\geq \frac{\rho \bi u^T S(\bi \beta)\bi u}{1+\rho \max_{1\leq i \leq n} \|\hat{\bi V}_i(\bi \beta)\|}-|\sum_{i=1}^n \omega_{ni}\bi u^T\hat{\bi V}_i(\bi \beta)|.
\end{align}
By Corollary A.1. of Jing, Yuan and Zhou (\cite{Jing2009}), $\max_{1\leq i \leq n} \|\hat{\bi V}_i(\bi \beta)\|=o_p(n^{1/2})$. From Lemma \ref{S}, 
$S(\bi \beta)=\bi \Sigma^{(\bi \beta_0)}_{r \times r}+o(1)$ uniformly for $\bi \beta \in B(\bi \beta_0, \delta_n)$, it follows that 
\begin{align}\label{order-lambda1}
\rho=\|\bi \lambda\|=O_p(n^{-1/2})+O_p(\|\bi \beta-\bi \beta_0\|).
\end{align}
Then we will show that $\bi \lambda-\bi \xi=O_p(n^{-1/2})$. 
Note that 
\begin{align*}
&\left | \sum_{i=1}^n \omega_{ni} \log (1+\bi \lambda^T \hat{\bi V}_i(\bi \beta))- \sum_{i=1}^n \omega_{ni} \log (1+\bi \lambda^T \bi g(X_i, \bi \beta))\right | \\
&=\left | \sum_{i=1}^n \omega_{ni} \frac{\bi \lambda^T \big (\hat{\bi V}_i(\bi \beta)-\bi g(X_i, \bi \beta)\big )}{1+\eta_i}\right |\\
&= \left | \sum_{i=1}^n \omega_{ni} \frac{\bi \lambda^T \bi R_{ni}(\bi \beta)}{1+\eta_i}\right |\\
&\xrightarrow{p} 0,
\end{align*}
where $|\eta_i| \leq \max_{1 \leq i \leq n} \left | \bi \lambda^T \hat{\bi V}_i(\beta)\right |+\max_{1 \leq i \leq n} \left | \bi \lambda^T \bi g(X_i, \bi \beta)\right |= o_p(1)$ by conditions \textbf{C1} and (\ref{order-lambda}). The last convergence comes from (\ref{hoeffding}) and (\ref{order-lambda}). Set $Q_{\bi \beta}(x, \bi \lambda)=\log (1+\bi \lambda^T \bi g(X_i, \bi \beta))$ in Theorem 4.1 of Wooldridge (\cite{Wooldridge1999}), along with conditions \textbf{C1, C2}, (\ref{order-lambda}) and the fact that $\log(1+x) < x$, we have $\bi \lambda-\bi \xi=O_p(n^{-1/2})$, uniformly for $\bi \beta \in B(\bi \beta_0, \delta_n)$.

Finally, 
let $\gamma_i(\bi \beta)=\bi \lambda^T\hat{\bi V}_i(\bi \beta)$. Then from (\ref{order-lambda1}) and Corollary A.1 of Jing, Yuan and Zhou (\cite{Jing2009}), 
\begin{align*}
\max_{1\leq i \leq n}|\gamma_i(\bi \beta)|=O_p(n^{-1/2})+O_p(\|\bi \beta-\bi \beta_0\|)) o(n^{1/2})=o_p(1)
\end{align*}
for all $\bi \beta$ in $B(\bi \beta_0, \delta_n)$ with $\delta_n=O_p(n^{-1/2})$.
Expanding (\ref{lambda}), 
\begin{align*}
&\bi 0=f(\bi \lambda)=\sum_{i=1}^n \omega_{ni}\hat{\bi V}_i(\bi \beta)\{1-\gamma_i(\bi\beta)+\frac{\gamma^2_i(\bi \beta)}{1+\gamma_i(\bi \beta)}\}\\
&=\sum_{i=1}^n\omega_{ni}\hat{\bi V}(\bi \beta)-S(\bi \beta) \bi \lambda+\sum_{i=1}^n\omega_{ni}\hat{\bi V}(\bi \beta)\frac{\gamma^2_i(\bi \beta)}{1+\gamma_i(\bi \beta)}
\end{align*}
where the last term is bounded by 
\begin{align*}
\sum_{i=1}^n \omega_{ni}\|\hat{\bi V}_i(\bi \beta)\|^3 \|\bi \lambda\|^2|1+\gamma_i(\bi \beta)|^{-1}=o(n^{1/2})O_p(n^{-1})O_p(1)=o_p(n^{-1/2})
\end{align*}
in which 
\begin{align} \label{order-v3}
\sum_{i=1}^n \omega_{ni}\|\hat{\bi V}(\bi \beta)\|^3=\|\hat{\bi V}(\bi \beta)\|\sum_{i=1}^n \omega_{ni}\hat{\bi V}(\bi \beta) \hat{\bi V}(\bi \beta)^T=o(n^{1/2}).
\end{align}

Therefore, $\bi \lambda=S^{-1}(\bi \beta)\sum_{i=1}^n \omega_{ni}\hat{\bi V}_i(\bi \beta)+o_p(n^{-1/2})$.

\begin{lemma}
Under conditions \textbf{C1-C4}, there exists a sequence $\{\delta_n\}$ and a constant $k>0$ such that $\Gamma(\bi \beta)\leq -k\| \bi \beta-\bi\beta_0\|^2$ for $\bi \beta \in B(\bi \beta_0, \delta_n)$. 
\end{lemma}
The proof can be found in Lemma 6 of Li, Xu and Zhou (\cite{Li2016}).

\begin{lemma}\label{decom1}
Under conditions \textbf{C1-C6}, we have
\begin{align*}
\Gamma_n(\bi \beta)=\Gamma(\bi \beta)+O_p(n^{-1/2} \|\bi \beta-\bi \beta_0\|)+o_p(\|\bi \beta-\bi \beta_0\|^2)+O_p(n^{-1})
\end{align*}
uniformly for all $\bi \beta \in B(\bi \beta_0, \delta_n)$.
\end{lemma}
\textbf{Proof.}
Note that
\begin{align}
&\sum_{i=1}^n\omega_{ni}\{\log(1+\bi \lambda^T \hat{\bi V}_i(\bi \beta))-\mathbb E[\log(1+\bi \xi^T \bi g(X_1, \bi \beta))]\}\nonumber \\
&=\sum_{i=1}^n\omega_{ni}\{\log(1+ \bi \lambda^T \hat{\bi V}_i(\bi \beta))-\log(1+\bi \xi^T \hat{\bi V}_i(\bi \beta))\}\nonumber \\
&+\sum_{i=1}^n\omega_{ni}\{\log(1+\bi \xi^T \hat{\bi V}_i(\bi \beta))-\log(1+\bi \xi^T \bi g(X_i, \bi \beta))\} \nonumber \\
&+\sum_{i=1}^n\omega_{ni}\{\log(1+\bi \xi^T  \bi g(X_i, \bi \beta))-\mathbb E[\log(1+\bi \xi^T \bi g(X_1, \bi \beta))]\}\nonumber\\
&:=A_1+A_2+A_3.\nonumber
\end{align}
\begin{align}
&A_1=\sum_{i=1}^n\omega_{ni}\{\log(1+\bi \lambda^T \hat{\bi V}_i(\bi \beta))-\log(1+\bi \xi^T \hat{\bi V}_i(\bi \beta))\}\nonumber \\
&=(\bi \lambda-\bi \xi)^T\sum_{i=1}^n\dfrac{\omega_{ni}\hat{\bi V}_i(\bi \beta)}{1+\bi \xi^T\hat{\bi V}_i(\bi \beta)}+
\frac{1}{2}(\bi \lambda-\bi \xi)^T\sum_{i=1}^n\dfrac{-\omega_{ni}\hat{\bi V}_i(\bi \beta)\hat{\bi V}_i(\bi \beta)^T}{(1+\bi \zeta^T\hat{\bi V}_i(\bi \beta))^2}( \bi \lambda-\bi \xi)\nonumber\\
&=O_p(n^{-1})+O_p(n^{-1/2})\| \bi\beta-\bi \beta_0\|),\nonumber
\end{align}
where each element of $ \bi\zeta_i$ lies between the corresponding one of $\bi \lambda$ and $\bi \xi$.
The last equality follows from Lemma \ref{S}, (\ref{weight-sum}) and Lemma \ref{||lambda||}.

\begin{align}
&A_2=\sum_{i=1}^n\omega_{ni}\{\log(1+\bi \xi^T \hat{\bi V}_i(\bi \beta))-\log(1+\bi \xi^T \bi  g(X_i, \bi \beta))\}\nonumber\\
&=\sum_{i=1}^n\omega_{ni}\{\frac{\bi \xi^T}{1+\eta_{1i}}(\hat{\bi V}_i(\bi \beta)-\bi g(X_i, \bi \beta))\}\nonumber\\
&=\sum_{i=1}^n\omega_{ni}\frac{\bi \xi^T \bi R_{ni}(\bi \beta)}{1+\eta_{1i}}\nonumber \\
&=O_p(1)\left\{O_p(n^{-1/2})+O_p(\|\bi \beta-\bi \beta_0\|)\right\}O_p(n^{-1/2})\nonumber \\
&=O_p(n^{-1/2}\|\bi \beta-\bi \beta_0\|))+O_p(n^{-1}),\nonumber
\end{align}
where $|\eta_{i1}| \leq \|\bi \xi \| \left \{ \max_{1 \leq i \leq n} \|\hat{\bi V}_{i}(\bi \beta) \|+  \max_{1 \leq i \leq n} \|\bi g(X_i, \bi \beta) \|\right \}=0_p(1)$.
\begin{align*}
&A_3=\sum_{i=1}^n\omega_{ni}\{\log(1+\bi \xi^T \bi g(X_i, \bi \beta))-\mathbb E[\log(1+\bi \xi^T \bi g(X_1,\bi \beta))]\}\nonumber\\
&=\sum_{i=1}^n\omega_{ni}\{\bi \xi^T \bi g(X_i, \bi \beta)-\mathbb E[\xi^T \bi g(X_1, \bi \beta))]\}\\
&-\frac{1}{2}\sum_{i=1}^n\omega_{ni}\{(\bi \xi^T \bi g(X_i, \bi \beta))^2-\mathbb E[\bi \xi^T \bi g(X_1, \bi \beta))]^2\}\nonumber\\
&+\frac{1}{3}\sum_{i=1}^n\omega_{ni}\{\frac{(\bi \xi^T \bi g(X_i, \bi \beta))^3}{(1+\eta_{2i})^3}-\mathbb E[\frac{(\bi \xi^T \bi g(X_1, \bi \beta))^3}{(1+\eta_2)^3}]\}\nonumber\\
&:=A_{31}+A_{32}+A_{33},
\end{align*}
where $|\eta_{21}| \leq \|\bi \xi \| \left \{ \max_{1 \leq i \leq n} \|\bi g(X_i, \bi \beta) \|\right \}=o_p(1)$ and $|\eta_{2}| \leq \|\bi \xi \|   \|\bi g(X_1, \bi \beta) \|=o_p(1).$

By Lemma \ref{||lambda||} and condition \textbf{C5}, we have 
\begin{align*}
&|A_{31}|=\left |\sum_{i=1}^n\omega_{ni}\{\bi \xi^T \bi g(X_i, \bi \beta)-\mathbb E[\bi \xi^T \bi g(X_1, \bi \beta))]\} \right | \\
&\leq \|\bi \xi\|\|\sum_{i=1}^n \omega_{ni}\left\{\bi  g(X_i, \bi \beta)-\mathbb E[ \bi g(X_1, \bi \beta))\right\}\|\\
&=\left\{O_p(n^{-1/2})+O_p(\|\bi \beta-\bi \beta_0\|)\right\} O_p(n^{-1/2})\\
&=O_p(n^{-1})+O_p(n^{-1/2}\|\bi\beta-\bi\beta_0\|),\\
&|A_{32}|=\left |\frac{1}{2}\sum_{i=1}^n\omega_{ni}\left \{(\bi \xi^T \bi g(X_i, \bi \beta))^2-\mathbb E[\bi \xi^T \bi g(X_1, \bi \beta))]^2 \right\} \right |\\
&\leq \frac{1}{2}\|\bi \xi\|^2\|\sum_{i=1}^n \omega_{ni}\left\{ \bi g(X_i, \bi\beta)^T \bi g(X_i, \bi \beta)-\mathbb E[ \bi g(X_1, \bi \beta)^T \bi g(X_1,\bi \beta)\right\}\|\\
&=o_p(n^{-1})+o_p(\|\bi \beta-\bi \beta_0\|^2), \\
\end{align*}
and
\begin{align*}
&|A_{33}|= \left |\frac{1}{3}\sum_{i=1}^n\omega_{ni} \left \{\frac{(\bi \xi^T \bi g(X_i, \bi \beta))^3}{(1+\eta_{2i})^3}-\mathbb E[\frac{(\bi \xi^T \bi g(X_1, \bi \beta))^3}{(1+\eta_2)^3}]\right \}\right |\\
&\leq \frac{1}{3}  \|\bi \xi\|^2\Big \|\sum_{i=1}^n \omega_{ni} \left \{\frac{\bi \xi^T \bi g(X_i, \bi \beta) \bi g(X_i, \bi \beta)^T \bi g(X_i, \bi \beta)\bi g(X_1, \bi \beta)^T  \bi g( X_1, \bi \beta)}{(1+\eta_{2i})^3}\right.\\
&\left.-\mathbb E\frac{\bi \xi^T \bi g(X_1, \bi \beta)\bi g(X_1, \bi \beta)^T\bi  g(X_1, \bi \beta)}{(1+\eta_2)^3}\right \} \Big \|\\
&=o_p(n^{-1})+o_p(\|\bi \beta-\bi \beta_0\|^2).\\
\end{align*}

\begin{lemma}\label{decom2}
Under conditions \textbf{C1-C6}, we have
\begin{align*}
\Gamma_n(\bi \beta)=\frac{1}{2}(\bi \kappa-\bi \kappa_0)^T \bi V(\bi \kappa-\bi \kappa_0)-\frac{1}{\sqrt{n}}(\bi \kappa-\bi \kappa_0)^T \bi Q_n+o_{p}(n^{-1})
\end{align*}
uniformly for $\bi \beta \in B(\bi \beta_0, \delta_n)$ with $\delta_n=O_p(n^{-1/2})$, here, $\bi \kappa=(\bi \lambda, \bi \beta)^T$, with true value $\bi \kappa_0=(0, \bi \beta_0)^T$, and

\begin{align*}
\bi Q_n=\left(\sqrt{n}\sum_{i=1}^n \omega_{ni} \hat{V}_{il}(\bi \beta_0), ..., \sqrt{n}\sum_{i=1}^n \omega_{ni}\hat{V}_{ir}(\bi \beta_0) ), \bi 0_q\right )^T,
\end{align*}
where $\bi 0_q$ is a vector of $q$ zeros.
\end{lemma}
\textbf{Proof.} Applying Taylor expansion to $\Gamma_n(\bi \beta)$, we have
\begin{align*}
&\Gamma_n(\bi \beta)=-\sum_{i=1}
^n\omega_{ni}\log(1+\bi \lambda^T\hat{\bi V}_i(\bi \beta))\\
&=-\sum_{i=1}^n\omega_{ni}\bi \lambda^T\hat{\bi V}_i(\bi \beta)+\frac{1}{2}\sum_{i=1}^n\omega_{ni}(\bi \lambda^T\hat{\bi V}_i(\bi \beta))^2-\frac{1}{3}\sum_{i=1}^n\omega_{ni}\frac{(\bi \lambda^T\hat{\bi V}_i(\bi \beta))^3}{(1+\eta_{3i})^3}\\
&:=J_1+J_2+J_3,
\end{align*}
where $|\eta_{3i}|=o_p(1)$ as before. 
As $\|\bi \lambda\|=O_p(n^{-1/2})$ and (\ref{order-v3}), we have
\begin{align}\label{J3}
J_3=O_p(n^{-3/2})O_p(n^{1/2})=o_p(n^{-1}).
\end{align}
For $J_1$, it can be rewritten as
\begin{align*}
&J_1=-\sum_{i=1}^n\omega_{ni}\bi \lambda^T\hat{\bi V}_i(\bi \beta_0)-\sum_{i=1}^n\omega_{ni}\bi \lambda^T[\hat{\bi V}_i(\bi \beta)-\hat{\bi V}_i(\bi \beta_0)]\\
&:=J_{11}+J_{12},
\end{align*}
where
\begin{align*}
&J_{11}=-\sum_{i=1}^n\omega_{ni}\bi \lambda^T\hat{\bi V}_i(\bi \beta_0)=-\bi \lambda^T [\sum_{i=1}^n\omega_{ni}\hat{\bi V}_i(\bi \beta_0)-\mathbb{E}\hat{\bi V}_i(\bi \beta_0)]\\
&=-\bi \lambda^T n^{-1/2} n^{1/2}\sum_{i=1}^n\omega_{ni}\hat{\bi V}_i(\bi \beta_0)-\mathbb{E}\hat{\bi V}_i(\bi \beta_0)\\
&= -n^{1/2} (\bi \kappa-\bi \kappa_0)^T \bi Q_n
\end{align*}
and 
\begin{align*}
&J_{12}=-\sum_{i=1}^n\omega_{ni}\bi \lambda^T[\hat{\bi V}_i(\bi \beta)-\hat{\bi V}_i(\bi \beta_0)]\\
&=-\bi \lambda^T\sum_{i=1}^n\omega_{ni}[\hat{\bi V}_i(\bi \beta)-\mathbb E \bi H(X_1, X_2; \bi \alpha_0, \bi \beta)-\hat{\bi V}_i(\bi \beta_0)+\mathbb E \bi H(X_1, X_2; \bi \alpha_0, \bi \beta_0)]\\
&-\bi \lambda^T \mathbb E \bi H(X_1, X_2; \bi \alpha_0, \bi\beta).\\
\end{align*}
By Lemma A.1 of Jiang $et$ $al.$ (\cite{Jiang2011}) and Lemma \ref{||lambda||}, the first term is
\begin{align*}
O_p(\sqrt{\sum_{i=1}^n \omega^2_{ni}}) \|\bi \lambda\|=O_p(\sqrt{\sum_{i=1}^n \omega^2_{ni}}) O_p(\sqrt{\sum_{i=1}^n \omega^2_{ni}})=o_p(\sum_{i=1}^n \omega^2_{ni})=o_p(n^{-1})
\end{align*}
where the last equality follows from (C6); for the second term, we have 
\begin{align*}
&-\bi \lambda^T \mathbb E \bi H(X_1, X_2; \bi \alpha_0, \bi \beta)\\
&=-\bi \lambda^T \frac{\partial}{\partial \bi \beta^T}\mathbb E \bi H(X_1, X_2;\bi \alpha_0, \bi\beta_0)(\bi\beta-\bi \beta_0)+o_p(n^{-1})\\
&=\bi \lambda^T V_{12} (\bi\beta-\bi \beta_0)+o_p(n^{-1}).
\end{align*}
Therefore, 
\begin{align}\label{J1}
J_1=-n^{-1/2} (\bi \kappa-\bi \kappa_0)^T \bi Q_n+\bi \lambda^T V_{12} (\bi \beta-\bi \beta_0)+o_p(n^{-1}).
\end{align}
Finally,
\begin{align*}
&J_2=\frac{1}{2} \sum_{i=1}^{n} \omega_{ni} [\bi \lambda^T \hat{\bi V}_i(\beta)]^2\\
&=\frac{1}{2} \sum_{i=1}^{n} \omega_{ni} [\bi \lambda^T \hat{\bi V}_i(\bi \beta_0)]^2+\frac{1}{2} \sum_{i=1}^{n} \omega_{ni} \{[\bi \lambda^T \hat{\bi V}_i(\bi \beta)]^2-[\bi \lambda^T \hat{\bi V}_i(\bi \beta_0)]^2\}\\
&:=J_{21}+J_{22},
\end{align*}
where 
\begin{align*}
&J_{21}=\frac{1}{2} \sum_{i=1}^{n} \omega_{ni} [\bi \lambda^T \hat{\bi V}_i(\bi \beta_0)]^2=\frac{1}{2} \bi \lambda^T[\sum_{i=1}^{n} \omega_{ni} \hat{\bi V}_i(\bi \beta_0) \hat{\bi V}^T_i(\bi \beta_0)-\bi \Sigma^{(\bi \beta_0)}_{r \times r}] \bi \lambda+\frac{1}{2}  \bi \lambda^T  \bi \Sigma^{(\bi \beta_0)}_{r \times r}  \bi \lambda\\
&=o_p(\|\bi \lambda\|^2)+\frac{1}{2} \bi \lambda^T  V_{11}   \bi \lambda=o_p(n^{-1})+\frac{1}{2} \bi \lambda^T  V_{11}   \bi \lambda,\\
&J_{22}=\frac{1}{2} \sum_{i=1}^{n} \omega_{ni} \left \{[\bi \lambda^T \hat{\bi V}_i(\bi \beta)]^2-[\bi \lambda^T \hat{\bi V}_i(\bi \beta_0)]^2 \right \}\\
&=\frac{1}{2} \bi  \lambda^T \sum_{i=1}^{n} \omega_{ni} [ \hat{\bi V}_i(\bi \beta)  \hat{\bi V}^{T}_i(\bi \beta)- \hat{\bi V}_i(\bi \beta_0)  \hat{\bi V}^{T}_i(\bi \beta_0)] \bi \lambda\\
&=\frac{1}{2}  \bi \lambda^T \left [\sum_{i=1}^{n} \omega_{ni}  \hat{\bi V}_i(\bi \beta)  \hat{\bi V}^{T}_i(\bi \beta)- \bi \Sigma^{(\bi \beta_0)}_{r \times r}- \sum_{i=1}^{n} \omega_{ni}\hat{\bi V}_i(\bi \beta_0)  \hat{\bi V}^{T}_i(\bi \beta_0)+ \bi\Sigma^{(\bi \beta_0)}_{r \times r} \right ]\bi \lambda\\
&=o_p(n^{-1}).
\end{align*}
Thus,
\begin{align}\label{J2}
J_2=\frac{1}{2} \bi \lambda^T V_{11}  \bi \lambda+o_p(n^{-1}).
\end{align}
Then the conclusion follows from (\ref{J3}),  (\ref{J1}) and  (\ref{J2}).

\begin{lemma}\label{decom3}
Under conditions \textbf{C1-C6}, we have
\begin{align*}
&\Gamma_n(\bi \beta)=-\frac{1}{2}(\bi \beta-\bi \beta_0)^T V^T_{12} V^{-1}_{11} V_{12}(\bi \beta-\bi \beta_0)+\frac{1}{\sqrt{n}}(\bi \beta-\bi\beta_0)^T V^{T}_{12} V^{-1}_{11}T_n\\
&-\frac{1}{2}n^{-1}T^{T}_nV^{-1}_{11}T_n +o_{p}(n^{-1})
\end{align*}
uniformly for $\bi \beta \in B(\bi \beta_0, \delta_n)$ with $\delta_n=O_p(n^{-1/2})$, and 
\begin{align*}
\tilde{\bi \beta}-\bi \beta_0=\frac{1}{\sqrt{n}} (V^{T}_{12}V^{-1}_{11}V_{12})^{-1}V^{T}_{12}V^{-1}_{11} T_n+o_p(n^{-1/2})
\end{align*}
\begin{align*}
T_n=\left(\sqrt{n}\sum_{i=1}^n \omega_{ni}\hat{ V}_{il}(\bi \beta_0), ..., \sqrt{n} \sum_{i=1}^n \omega_{ni} \hat{V}_{ir}(\bi \beta_0) )\right )^T.
\end{align*}
\end{lemma}
\textbf{Proof.}
By Lemma \ref{||lambda||}, we have $\bi \lambda=V^{-1}_{11}[n^{-1/2} T_n-V_{12}(\bi \beta-\bi \beta_0)]+o_{p}(n^{-1/2})$ uniformly for $\bi \beta \in B(\bi \beta_0, \delta_n)$ with $\delta_n=O_p(n^{-1/2})$. Therefore,
\begin{align*}
&\Gamma_n(\bi \beta)=\frac{1}{2}\bi \lambda^T V_{11} \bi \lambda+\bi \lambda^T V_{12} (\bi \beta-\bi \beta_0)-\frac{1}{\sqrt{n}}\bi \lambda^T T_n+o_{p}(n^{-1})\\
&=-\frac{1}{2}(\bi \beta- \bi \beta_0)^T V^{T}_{12} V^{-1}_{11}V_{12}(\bi \beta-\bi \beta_0)+\frac{1}{\sqrt{n}}(\bi \beta-\bi \beta_0)^T V^{T}_{12} V^{-1}_{11}T_n\\
&-\frac{1}{2}n^{-1} T^{T}_n V^{-1}_{11}T_n+o_{p}(n^{-1}).
\end{align*}
By Theorem 1 of Wooldridge (\cite{Wooldridge1999}), we have  $\bi \beta-\bi \beta_0=O_{p}(n^{-1/2})$. Moreover, from Theorem 2 of \cite{Wooldridge1999},
we have
\begin{align*}
\tilde{\bi \beta}-\bi \beta_0=\frac{1}{\sqrt{n}} (V^{T}_{12}V^{-1}_{11}V_{12})^{-1}V^{T}_{12}V^{-1}_{11} T_n+o_p(n^{-1/2}).
\end{align*}

\noindent\textbf{PROOF OF THEOREM \ref{aschinuisance}.} Denote $V_{22.1}=-V^{T}_{12}V^{-1}_{11}V_{12}$,  by Lemma \ref{decom3},
\begin{align*}
&\Gamma_n( \tilde{\bi \beta})=-\frac{1}{2}n^{-1}   T^{T}_n  V^{-1}_{11}V_{12} V^{-1}_{22.1}V^{T}_{12} V^{-1}_{11} T_n
-\frac{1}{2}n^{-1}  T^{T}_n  V^{-1}_{11}  T_n+o_{p}(n^{-1})\\
&=-\frac{1}{2}n^{-1}  T^{T}_n  V^{-1/2}_{11} D V^{-1/2}_{11}  T_n+o_{p}(n^{-1}),
\end{align*}
where $D=V^{-1/2}_{11} \{I+ V_{12} V^{-1}_{22.1}V^{T}_{12} V^{-1}_{11}\}V^{1/2}_{11}$,or, 
\begin{align*}
l(\bi \alpha_0, \tilde{\bi \beta})=T^{T}_n  V^{-1/2}_{11} D V^{-1/2}_{11} T_n+o_{p}(1).
\end{align*}
Applying Lemma 1 of Jiang $et$ $al.$cite{Jiang2011} and Slutsky's theorem, we have 
\begin{align*}
V^{-1/2}_{11} T_n=V^{-1/2}_{11} n^{1/2}\sum_{i=1}^n \omega_{ni} \hat{\bi V_i}(\bi \beta_0) \stackrel{d}{\rightarrow} \sqrt{c} N_{r}(\bi 0, I_{r \times r})
\end{align*}
since  $\lim_{n \to \infty} \sum_{i=1}^n n\omega^2_{ni}=c$.
Similar to Molanes Lopez, Van Keilegom and Veraverbeke (\cite{Molanes2009}), it can be checked that $D$ is symmetric, idempotent and with trace $p$. Thus, $l(\bi \beta_0)\stackrel{d}{\rightarrow} c \chi^2_p$.

\par

\end{document}